\documentclass[nonacm]{acmart}

\usepackage{tikz}
\usetikzlibrary{decorations.pathreplacing}

\AtBeginDocument{%
  }

\begin{document}

\title{The Design of Informative Take-Over Requests for Semi-Autonomous Cyber-Physical Systems: Combining Spoken Language and Visual Icons in a Drone-Controller Setting}

\author{Ashwini Gundappa}
\author{Emilia Ellsiepen}
\authornotemark[1]
\email{emilia@lst.uni-saarland.de}
\affiliation{%
 \institution{Saarland University}
 \country{Germany}
}
\author{Lukas Schmitz}
\author{Frederik Wiehr}
\affiliation{%
 \institution{DFKI}
  \country{Germany}
}
\author{Vera Demberg}
\affiliation{%
\institution{Saarland University}
 \country{Germany}
}

\begin{abstract}
The question of how cyber-physical systems should interact with human partners that can take over control or exert oversight is becoming more pressing, as these systems are deployed for an ever larger range of tasks. 
 Drawing on the literatures on handing over control during semi-autonomous driving and human-robot interaction, we propose a design of a take-over request that combines an abstract pre-alert with an informative TOR: Relevant sensor information is highlighted on the controller's display, while a spoken message verbalizes the reason for the TOR. We conduct our study in the context of a semi-autonomous drone control scenario as our testbed.
 The goal of our online study is to assess in more detail what form a language-based TOR should take. Specifically, we compare a full sentence condition to shorter fragments, and test whether the visual highlighting should be done synchronously or asynchronously with the speech. 
 Participants showed a higher accuracy in choosing the correct solution with our bi-modal TOR and felt that they were better able to recognize the critical situation. 
 Using only fragments in the spoken message rather than full sentences did not lead to improved accuracy or faster reactions. Also, synchronizing the visual highlighting with the spoken message did not result in better accuracy and response times were even increased in this condition.
\end{abstract}



\maketitle
\section*{Acknowledgements}
This work was supported by the DFG in grant 389792660 as part of TRR 248 (https://perspicuous-computing.science).

\section{Introduction}\label{sec:intro}

Much effort is currently going into the development of autonomous cars and other (semi-)autonomous cyber-physical systems. These systems carry the promise to reduce human error, increase safety and efficiency, provide comfort, and reduce humans' responsibilities \cite{Hassenzahl}.
However, many of these systems currently cannot be operated in a fully autonomous mode, i.e.~they may need to ask humans for assistance on decisions in unexpected
situations or system failure.
Furthermore, even fully autonomous AI systems are required to allow for effective oversight by a human, if their decisions are possibly consequential: see the AI act
\footnote{\url{https://digital-strategy.ec.europa.eu/en/policies/european-approach-artificial-intelligence}}.

In order for the take-over or oversight process to be successful, it is necessary to ensure that the human partner, who might not be fully aware of the state of the cyber-physical system, can be effectively informed of the reason for a request to take over control and any imminent dangers.
In this paper, we investigate how the take-over messages should best be designed in terms of an effective conveyance of critical information from the system to the controller.

In the case of a take-over request (TOR), the human has to assess the information made available by the system interface and decide how to resolve the situation. In order for a take-over to be effective, it is important for the system to draw the human's attention to the information that led to the critical situation \cite{E2:endsley1996automation,10.1145/3122986.3123020}.
In autonomous driving, which is the most well-researched area within TOR design, it has been found to be beneficial to use alerting sounds as well as visual highlighting for important relevant information. Furthermore, spoken language has been found to be an effective modality for TORs which include the reason for take-over \cite{Politis2015} or post-hoc explanations \cite{Krber2018}. 
These findings lead us to \textbf{Hypothesis H1: a combination of visual highlighting of relevant sensor data on the controller display together with a spoken message mentioning the sensor information that caused the criticality will lead to a better performance than alerting the controller with an abstract sound signal or with visual highlighting alone.}

While existing studies have compared spoken language cues to other modalities or no cues, there is little work on the form of the spoken message itself. Our baseline is a message in the form of complete sentences that verbalizes the situation-relevant sensor information, but nothing more. An important consideration, however, is not to overload the human controller with information, as it might severely impact their ability to determine what to attend to and how to resolve the situation \cite{Koo}. Also, it is important for hand-over messages to be short in time-critical situations. Further insight on this can be gained by comparing the full sentences to an abbreviated message in the form of fragments that might introduce a lower cognitive load, and be faster to convey. In our study, we manipulate the form of the spoken message and propose \textbf{Hypothesis H2: Fragments will be faster (and possibly easier) to process than full sentences, and will therefore lead to better performance on the part of the human controller.}

Existing studies in autonomous driving have already shown that combining different modalities for signaling can lead to a faster reaction compared to single modality signals \cite{Politis2013}. In our study, the same pieces of sensory information will be presented in the visual and auditory (language) modality. A follow-up question for whether the combination of two modalities is beneficial is how exactly the combination is best implemented. To this end, we are going to compare two implementations: a simple addition, where the visual highlighting is presented for the whole time of the TOR, and a synchronized version, where sensor data is only highlighted when it is also mentioned in the spoken message. If the reason for multi-modal cues being more effective lies in a facilitated uptake of information, a synchronized version should be even more effective. If, on the other hand, the benefit is due to different modalities being preferred by different human controllers or in different situations, we will not see an improvement if the two are synchronized. This thus leads us to \textbf{Hypothesis H3: Performance will be improved when icons are highlighted synchronously with the mention of the critical information in the spoken utterance.}


As a testbed for our study, we use the drone control scenario. In this scenario, semi-autonomous drones operate autonomously most of the time, but may run into critical situations, for which control then needs to be handed over to a human operator. We chose the drone domain because 
we expect language to be even more important in the drone controller setting than in the more common setting of autonomous cars, since the amount of information to be shared is often more complex and so is the space of solutions. Also, in contrast to operating a car, where formal training is required, drones may be operated by less trained humans potentially lacking experience with specific situations. Here, conveying information is of increased importance and the form of a take-over request should be easy to process, even for naive users.  As a third motivation, it is a domain of increasing real-world importance, i.e., companies like Amazon with PrimeAir and Wing recently received official certification to start drone delivery services. 
And the deployment of drones has generated a broad range of applications including disaster relief \cite{Bravo2015}, monitoring \cite{Flammini}, surveillance \cite{Kaleem} and delivery services in heterogeneous fields like agriculture \cite{Ahirwar}, defense \cite{Coffey}, transport \cite{Khosravi}, industry \cite{Shahmoradi} and commerce \cite{Maghazei}.

\begin{figure}
    \centering
    \includegraphics[width=3in]{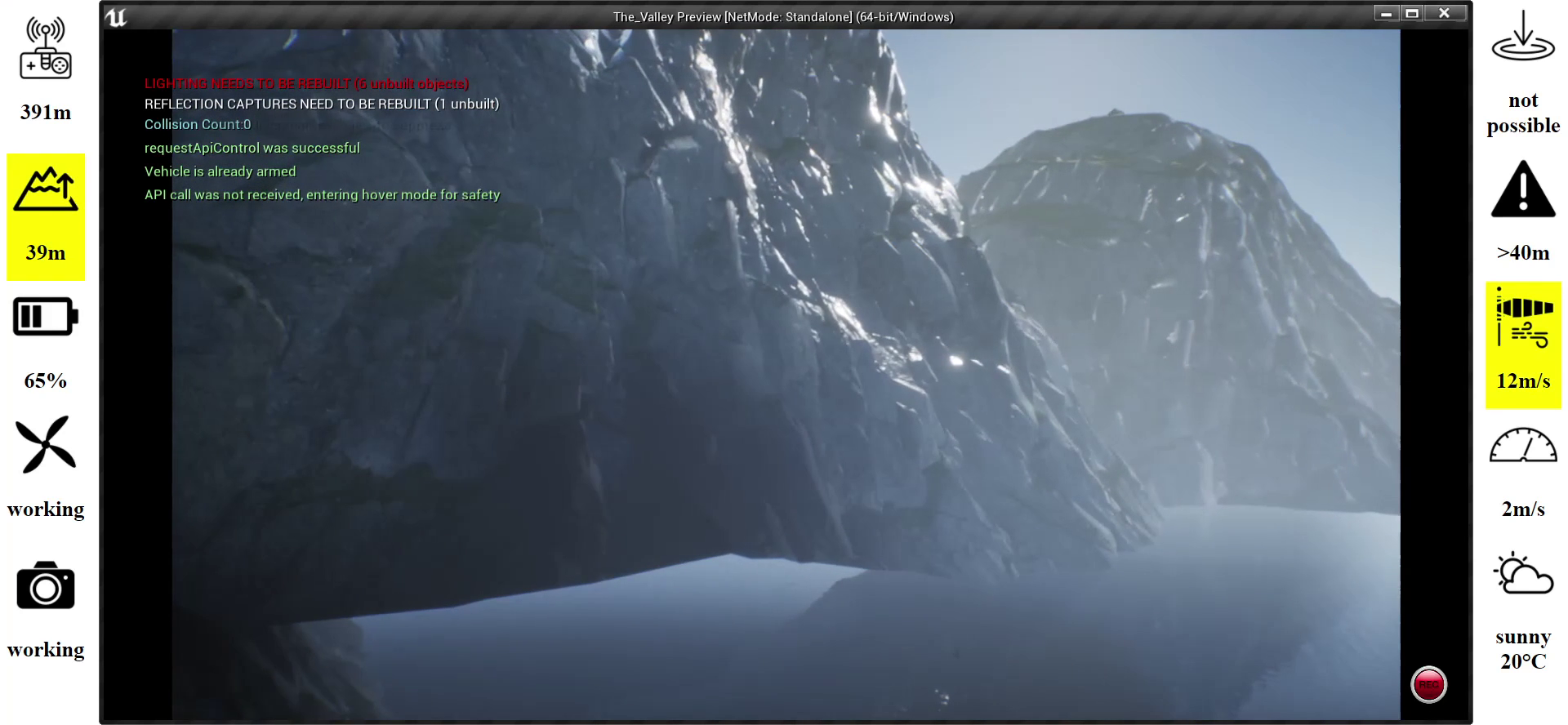}
    \caption{Example of a drone controller's display}
    \label{fig:scenario1}
\end{figure}

To illustrate our setting more concretely, consider the following situation from our study: The drone system becomes unstable due to strong winds at high altitude. While the sensors have gathered this information and the system therefore has initiated a take-over operation, the reason for take-over might not be obvious to the controller just looking at the drone's camera transmission. An instance for the controller's display is given in Figure \ref{fig:scenario1}: Here, the relevant icons (altitude and wind speed) are already highlighted, and at this point the controller would hear a spoken language message also pointing out the altitude and strong winds. Without these cues, the controller might try to avoid the mountain by going up, instead of reducing the altitude to avoid the winds, when prompted for a take-over.

In the remainder of this article, we introduce the related work and provide more background on modality, situation awareness, and timing in Section 2. The proposed designs for a TOR in the context of the mixed-initiative drone setting are described in Section 3. In Section 4, we evaluate the approach with a user study in which we compare our TOR designs. In Sections 5 and 6, we discuss the findings of our study and make recommendations for future research.

\section{Related Work}

While we are not aware of any prior work on the design of hand-over requests in our specific testbed, semi-autonomous drones, there is a multitude of user studies in the field of semi-autonomous driving as well as some related work in the field of aviation and human-robot interaction. Studies have focused on comparing different modalities and combinations thereof, as well as the influence of timing and situational awareness, and have used different measures such as reaction times to TORs, hand-over accuracy, and also subjective ratings of user experience. In the following, we will review these studies, pointing out how they are relevant to either the general design of a TOR in the drone scenario, or to the specific research questions H1 to H3.


\subsection{Situational Awareness and Providing Explanations}
When taking over from a semi-autonomous system, it is vital that the controller quickly regains situational awareness to react adequately, while a lack of the same can lead to accidents \cite{E2:endsley1996automation,E2:Adams1995}. Situation awareness describes the human's awareness of the
environment \cite{E2:endsley1996automation}, including awareness of
critical information for a task at hand, mainly known from aviation but also applied to semi-autonomous driving \cite{10.1145/3122986.3123020}. One way to induce situational awareness in the context of a hand-over is to provide an explanation along with the TOR.
State-of-the-art systems often lack the technical support for generating explanations early on in the handover procedure. Wiehr et al. \cite{WiehrICMI2021} present an approach to incorporate explanations in autonomous systems such as  drones and cars. The system architecture features an ontology which contains knowledge about the environment and the system's state. A description logics component is able to derive conclusions about situations classifying the situation as potentially critical. 

In the context of a take-over in the driving scenario, Wiehr et al. \cite{WiehrICMI2021}, contrasted the contribution of more time and more information: In their study, participants overall preferred systems that gave early warnings accompanied by an explanation, while a system giving the explanation at the same time as the TOR was not as well-liked as a system giving no explanation, but an early warning. Even when providing information post-hoc, it can positively influence future driving behavior, as trust in and understanding of the system are increased \cite{Krber2018, E2:pflegingetal17}.
Other studies showed that an auditory alert ahead of time caused participants to look more at the road before the handover and disengage from the secondary task earlier \cite{Heiden2017}. Also, giving early additional audio-visual explanations about an appropriate upcoming maneuver leads to faster responses and longer time to collision with obstacles \cite{Borojeni2018}. 

While explanations are thus beneficial, the user needs enough time to process them and should not be overloaded with information \cite{Koo}. A survey with professional pilots concluded that minimizing the exposure to unnecessary alerts, categorizing and prioritizing should be a general requirement for designing systems in aviation \cite{Veitengruber1978}.
Also, multiple signals require prioritization, for which the operators are often not trained~\cite{Bliss1995}, and this can cause accidents~\cite{Bliss2003}. 



\subsection{Timing of the Take-over}
Users might not be able to immediately take control and shift their attention to the primary tasks as they must disengage themselves from other tasks \cite{Hancock2017}. In the driving scenario, van der Heiden et al. \cite{vanderHeiden} investigated the influence of auditory pre-alerts in a dual-task scenario, which occurred 20 seconds before the TOR. They found that participants responded faster (braking) and showed an overall less elevated heart rate, indicating that users benefit from pre-alerts. In contrast, Walch et al., 2015 \cite{Walch} varied the timing between the TOR and the critical situation (fog) between 4 and 6 seconds, with video watching as a secondary task. They found no significant difference, suggesting that the participants reacted to the TOR immediately when it was made, even without assessing the situation. In this study, the response (grabbing the steering wheel) was significantly delayed if an explanation was presented in parallel to the TOR. It is worth noting here, however, that no situational awareness was required for the participant to respond. Gold et al. \cite{Gold2013} found shorter reaction times with shorter take-over times (7 vs. 5 seconds). Crucially, they also found that driving performance was better for the longer take-over time, thus suggesting an overall beneficial effect of a longer take-over. Focusing on drivers' preferences, Wiehr et al. \cite{Wiehr2021} found that a system with an explanation given 30 seconds prior to the critical situation and the actual TOR 15 seconds before it, is generally preferred to systems with no early alert and also to one where the explanation is given at the same time as the TOR.

In summary, the literature suggests that in general, drivers benefit from pre-alerts before the actual take-over, but should have time to process explanations before they need to take over control. These findings are expected to carry over to other HRI domains, in particular the drone scenario, with the additional complication that in general more sources of information need to be processed before taking over control. A prerequisite for allowing long take-over times is of course that the system detects a critical situation early on, which can be achieved based on information from the drone's sensors using description logic, if an ontology expressing criticality in the domain is available; see for instance \cite{ChangLREC2022}.

\subsection{Choice of Modality}

Multimodal interaction has
the potential to increase the usability and the safety of
operation \cite{E2:cohen2004tangible}. It has been used for mobile
applications and environments, including gesture and speech \cite{E2:wasinger2005integrating}, eye
tracking and face detection \cite{E2:muller2009reflectivesigns}, gaze-based  \cite{Walch2019, Schmidt2019} and tangible interaction \cite{E2:kalnikaite2011nudge} to adapt to the user's needs. 

In the context of human-robot interaction, multimodal input has been introduced in order for interactions to  be more failsafe, i.e. using speech, gaze, and tactile to increase the redundancy in the communication channel or the loss of a single modality \cite{bannat2009multimodal}. 
For the specific task of the transfer of a physical object between robot and human, gaze, gesture, and body posture were suggested to be important for a success of the operation in analogy to interactions between two humans \cite{Strabala2013, huber2009, moon2014}. 

As for our specific focus, the take-over of control between CPS and human, there are valuable insights in the use of different modalities from the driving setting.
Naujoks et al. \cite{Naujoks2014} contrasted purely visual warnings (flashing icons on the screen) with a bimodal warning, which additionally included an abstract audio tone. The combined warning resulted in significantly reduced reaction times and better lane-keeping in participants. In addition, a study looking at warnings of varying urgency found that purely auditory cues are harder to interpret, if not accompanied by a visual cue \cite{PolitisBeep}. Both modalities thus seem to play an important role, especially when the participant is required to interpret the cue. The visual modality can also  be implemented in the form of peripheral lights \cite{Borojeni2016} instead of a centrally presented visual cue which blocks vision of the road: In the drone scenario investigated here, visual cues will be presented in the margins around the camera view of the drone. 
In addition to faster and more accurate reactions, users were in general found to prefer multimodal signals, although annoyance also increased with an increase in modalities, especially when including tactile as a third modality \cite{Politis2015,Politis2013}. All of the above suggest that the combination of visual and auditory TOR will be the best in terms of performance and likability.

Some of the above work has focused purely on signaling take-over, using the same cue for all kinds of take-over situations \cite{Naujoks2014}, or solely encoding the urgency level \cite{Politis2013}. In these studies, the integration and synchronization of the different modalities appears not to be of interest, as they can easily be perceived in parallel. But even studies with more complex cues, e.g. Politis et al. \cite{Politis2015}, did not directly investigate how the modalities are integrated, or experiment with the synchronization of the different modalities, but presented all information in parallel.
In the drone scenario, different pieces of sensor information have to be processed (e.g. wind speed and altitude) in order to react adequately to the critical situation. 
In the visual modality, even if all information is given at once (i.e. highlighting both icons simultaneously), the user still has to fixate them one after the other to take up the information, while in the auditory modality, only one ordering can be instantiated and will be determined by the system, rather than the user (i.e. first mention strong winds, then mention the high altitude, or vice versa). Drawing from literature in psycholinguistics, we know that visual attention is closely coupled with language processing, i.e., objects that are mentioned in spoken language are quickly fixated \cite{Allopenna98} and visual information, in turn, can be used to facilitate language processing \cite{Tanenhaus95}. An open question in this context is whether synchronization of the visual and auditory modalities will facilitate the processing of the information given in the TOR and lead to better hand-over performance.

\section{Design of a take-over request}

The review on the literature from semi-autonomous driving and other HRI settings in the last section provides us with a number of guidelines to design a take-over request for a CPS in general and for our testbed of semi-autonoumous drones in particular: Firstly, a combination of modalities is expected to be more effective than just using one modality. Secondly, providing information that will induce situational awareness is expected to enhance the take-over. Thirdly, the controller needs time to process additional information and, more generally, pre-alerts before the take-over are beneficial. The proposed take-over request here takes into account these requirements by providing information before the need to take over, in both the visual and the auditory modality. 

Our setup is as follows: The controller who should handle the TOR does not have direct visual contact with the drone, but is provided with a visual display showing the drone's camera transmission, as well as sensor information in the form of visual icons. In the bi-modal TOR that we propose, relevant icons will be highlighted and a spoken message will verbalize the relevant sensor information. While previous research looked at how different types of audio information affect performance \cite{Koo}, or the importance of providing additional information \cite{Krber2018}, none of them looked at how short vs.~longer versions of the same information affect performance. To bridge this gap,  we vary the linguistic form of the spoken message as detailed below. Furthermore, studies investigating the employment of more than one modality have all presented the signal in parallel, while we propose to synchronize auditory with visual information in order to enhance information processing on the part of the controller.

\subsection{Visual Icons}
As we were not able to find drone-related icons in the literature, we constructed them based on the environment, the task of the drone, and the types of sensor information presumably available in a drone system. The icons used in the study include ones for distance to the remote control, altitude, and battery status; propeller and camera icons that indicate whether they are working or damaged; an icon indicating whether landing in the current location is possible, and icons indicating distance to the next physical obstacle, wind speed, drone speed, and finally weather.
As can be seen in Figure \ref{fig:scenario1}, the icons themselves are visual representations of the kind of information (e.g. a mountain with an upwards arrow for altitude), while the actual sensor output is displayed below (e.g. 39 m). For some icons, the appearance of the icon changes when sensor information changes (battery level, weather). All icons are displayed in the margins of the camera transmission so as not to disrupt the view of the transmission. When used in the TOR, an icon is highlighted in yellow.

\subsection{Message}
The messages define elements that lead to a critical situation, which can be more than one element in some cases. Ordering of elements in the messages is such that information which is more important to the decision comes first. If the elements cannot be ordered according to their importance, elements that are difficult to perceive come first. All messages are vocalized in English using the Google Text-to-Speech synthesizer with a male voice, as they can be logically balanced with tone and sound very natural. Each message is preceded by an audio beep, followed by approximately 1.5 seconds of silence.
In addition to the default version consisting of full sentences, we constructed messages just containing the vital parts of information in fragments as given in Table \ref{tab:message}. The complete audio signal (beep + silence+ spoken message) lasts between 4 to 6 seconds for sentences and between 3 to 4 seconds for fragments. 
\begin{table}[]
    \centering
    \begin{tabular}{l l l }
    \hline
    condition & speech message & length of speech signal\\\hline
     sentence    & The drone is flying in strong wind at a very high altitude. & 3.2 s \\
     fragment    & Strong wind! High altitude! & 2.1 s\\\hline
    \end{tabular}
    \caption{Example messages for TOR}
    \label{tab:message}
\end{table}

\subsection{Timing}
\begin{figure}
    \centering
\begin{tikzpicture}[decoration=brace,scale=.6]
    \draw [fill=yellow, yellow] (4,0) rectangle (14,1) ;  
    \draw [fill=yellow, yellow] (4,1) rectangle (14,2) ;  

    \draw[thick,->](0,0)--(23,0);
    \foreach \x/\xtext in {4/,14/, 21/}
          \draw(\x,3)--(\x,0) node[below] {\xtext};
    \node[align=center] at (4.2,4) {beep};
    \draw [thick,->] (4.2,3.8) -- (4.2,3.15);
    \node[right] at (5.3,4) {\textit{The drone is flying in strong wind at a very high altitude.}};
    \draw [thick,->] (6,3.8) -- (6,3.15);
    \node[right,align=center] at (0,2) {Autonomous\\mode};
    \node[right,align=center] at (21,2) {``Crash''};
        \node[align=center] at (17.5,2) {Take-over\\time};

   \node[above,inner sep=0pt] (sound) at (9,2)
    {\includegraphics[width=.39\textwidth]{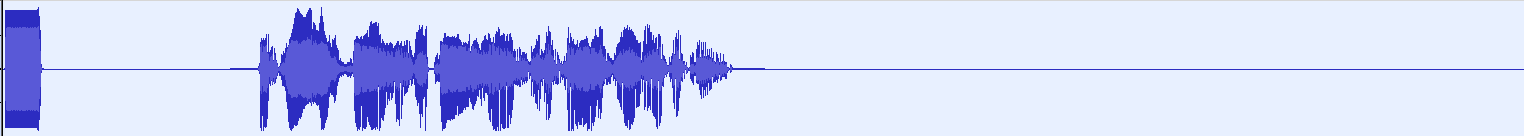}};     
   \node[above,inner sep=0pt] (alt) at (4.5,0)
    {\includegraphics[width=.04\textwidth]{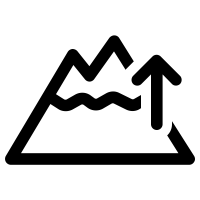}};    
   \node[above,inner sep=0pt] (wind) at (4.5,1)
    {\includegraphics[width=.04\textwidth]{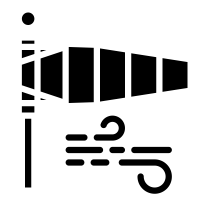}};   

  \end{tikzpicture}
  
  \begin{tikzpicture}[decoration=brace,scale=.6]
    \draw [fill=yellow, yellow] (7.9,0) rectangle (14,1) ;  
    \draw [fill=yellow, yellow] (7.1,1) rectangle (14,2) ;  

    \draw[thick,->](0,0)--(23,0);
    \foreach \x/\xtext in {4/-17,14/-7, 21/0}
          \draw(\x,3)--(\x,-8pt) node[below] {\xtext};

    \node[right,align=center] at (0,2) {Autonomous\\mode};
    \node[right,align=center] at (21,2) {``Crash''};
        \node[align=center] at (17.5,2) {Take-over\\time};

   \node[above,inner sep=0pt] (sound) at (9,2)
    {\includegraphics[width=.39\textwidth]{figures/sound}};     
   \node[above,inner sep=0pt] (alt) at (8.4,0)
    {\includegraphics[width=.04\textwidth]{figures/altitude.png}};    
   \node[above,inner sep=0pt] (wind) at (7.5,1)
    {\includegraphics[width=.04\textwidth]{figures/wind}};   
 \draw [thick,decorate,decoration={brace,amplitude=6pt,raise=0pt,mirror}] (4,-1) -- (14,-1);
\node[below,align=center] at (9,-1.2) {Take-over request (10 sec)};   
    \foreach \x in {1,2,...,21}  \draw (\x,0) -- (\x,-3pt);
  \end{tikzpicture}
    \caption{Take-over Process. Top: Version 3 with asynchronous visual highlighting and spoken message (sentence). Bottom: Version 5 with synchronous visual highlighting and spoken message (sentence).}
\label{fig:Process of TOR}
\end{figure}

As previously stated, auditory pre-alerts that occur prior to TOR result in better performance, and users benefit from early explanations \citep{vanderHeiden, Wiehr2021}. Also, with a shorter take-over time, the response may be faster but less accurate, resulting in poor reactions \citep{Gold2013}. We take this into account by allowing for a total take-over time of 17 seconds, which is further divided into 10 seconds for the TOR and a maximum of 7 seconds for the controller to react before the critical situation culminates, as shown in Figure \ref{fig:Process of TOR}. In the first part, an audio signal (beep) is used to grab the controller's attention. It is then followed by a message verbalizing the relevant sensor information that led to the request for   take-over. The same information is provided in the visual modality, by highlighting relevant icons starting either at the same time as the beep, or from their mentioning in the spoken message until take-over has finished. The second part (take-over time) is separated from the TOR, such that the controller does not have to process any more information when making a decision on how to react, as this has been shown to be problematic \cite{Wiehr2021, Walch}. Fixing it to 7 seconds is a compromise to give the controller enough time to come to a viable solution, while on the other hand forcing them to resolve the situation quickly to avoid unnecessary damage to the drone. We evaluate the appropriateness of the timing in the user study.

\subsection{Different versions of the TOR}\label{sec:TORdesign}

The TOR is presented in five different ways:
\begin{enumerate}
    \item \textbf{Baseline}: Two audio signals (beeps) will indicate when to take control. After the first beep, the controller has the opportunity to assess the situation with the help of the information displayed (visual icons); after the second beep the video ends and they have to select one of the options displayed.
    \item \textbf{Visual only}: An audio beep will indicate when to take control. After the beep, the controller has the opportunity to assess the situation with the help of the highlighted visual icons. After the video ends, they have to select one of the options displayed.
    \item \textbf{Language (sentences) + Visual with asynchronous highlighting}: An audio beep will indicate when to take control. After the beep, a spoken message explains the reason for the take-over in full sentences and participants have the opportunity to assess the situation with the help of the visual icons which are highlighted throughout the take-over request.
     \item \textbf{Language (fragments) + Visual with asynchronous highlighting}: An audio beep will indicate when to take control. After the beep, a spoken message explains the reason for the take-over in sentence fragments and participants have the opportunity to assess the situation with the help of the visual icons which are highlighted throughout the take-over request.
    \item \textbf{Language (sentences) + Visual with synchronous highlighting}: An audio beep will indicate when to take control. After the beep, participants have the opportunity to assess the situation with the help of a spoken message explaining the reason for the take-over in full sentences as well as the visual icons, which are highlighted when they are first mentioned and remain highlighted until the end of the take-over request.
\end{enumerate}

Versions 1 and 2 serve as baselines in the user study presented in \ref{sec:study}, while version 3 incorporates prior findings on TOR design in the driving setting as described above. Version 4 takes into account a possible benefit from shortening the linguistic message, while version 5 additionally shows a synchronization of visual and auditory information in order to improve processing.

\section{User Study} \label{sec:study}
The aim of our user study was to test hypotheses H1 to H3 in a setting with naive participants with a challenging task in the domain of HRI. The experiment is not trying to approximate an existing set-up of a drone controller in real life, but rather investigates the impact of the above-described manipulations of the TOR on performance and user experience measures in this illustrative setting.
For each of these hypotheses, we formulate more specific predictions in terms of accuracy of the decision, response time and user experience.

\subsection{Apparatus}


\begin{figure}
  \centering
    \includegraphics[width=0.5\textwidth]{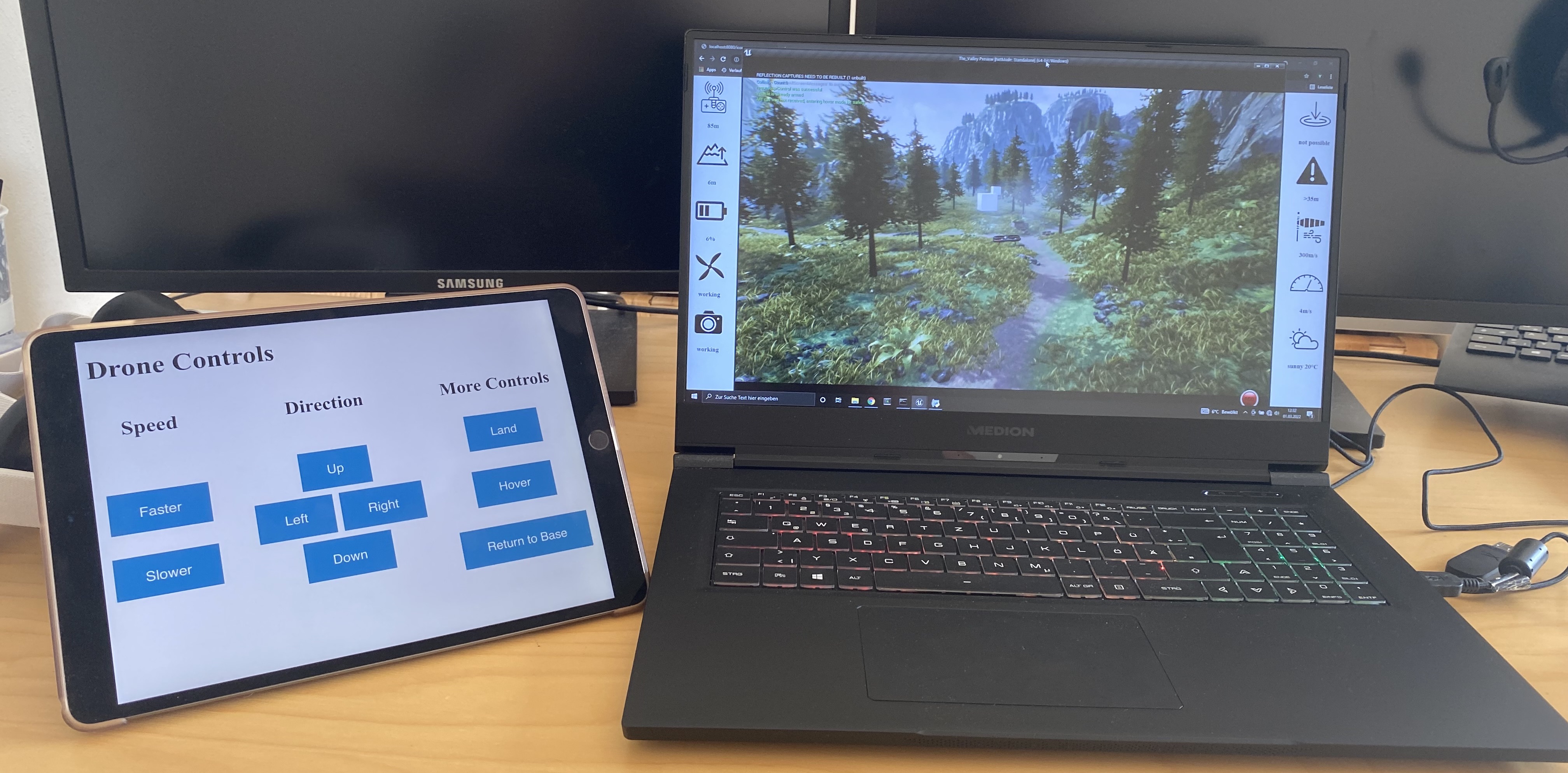}
    \caption{Autonomous drone simulator with explanations: secondary device (tablet) can be used to make a decision in a critical situation}
    \label{fig:fullsetup}
\end{figure}

In this work, we applied our simulation framework, which enables researchers to create mixed-initiative handover simulations using multimodal explanations for autonomous drones and highly automated driving. The software features  environmental variables which can be adjusted to create a variety of handover situations to perform studies and was used to create the scenarios/screenshots for the online survey. 
It is based Microsoft AirSim \cite{shah2018airsim} and  extended to be used with a secondary device (tablet) to make discrete decisions especially for this study. This allowed us to create the critical situation handovers with generated fragments  or full sentences depending on the condition in the online study (see Figure~\ref{fig:fullsetup}).

\subsection{Materials}
The objective in creating different scenarios was to have the possibility to get multiple measurements from the same participant, as well as to prevent unintended idiosyncrasies of one scenario having an overly large impact on the overall results. We thus introduced variability into the study to be able to generalize to other situations. Scenarios were intended to be of similar criticality and to be different enough to prevent strong learning effects over the course of the experiment. In addition, as we used accuracy as our main outcome, scenarios were designed to be solved non-trivially, as we can not test for experimental manipulations in the presence of ceiling effects. In order to achieve this, scenarios were created by varying the combination of problems that led to the critical situation and elements of the drone environment simulation. 

To identify relevant problems in drone-piloting settings, a questionnaire was administered to 10 drone users. The participants here mentioned different weather conditions, as well as obstacles and battery loss as common causes for hand-over or difficulties in drone piloting. Based on that, we created problem combinations of type  (technical problem, weather), (technical problem, obstacle) and (weather, obstacle).  Table ~\ref{table:1} shows the elements used to create scenarios. The number in brackets gives the total number of scenarios that contain the respective problem. 


In total, 20 different scenarios were created using two existing packages (i.e, park and valley environment, 10 scenarios each) which provide a large number of park props, trees, playgrounds, roads, mountains, sea, etc. In each scenario, the drone initially flew in autonomous mode in a predefined path for between 20 to 30 seconds. The drone then encountered a critical situation where it needed human intervention and it alerted the user by sending a take-over request (TOR). The five types of TOR described in the last section were instantiated for each scenario and a video recording for each of these five versions was created. The videos lasted exactly until the end of the TOR, i.e.~10 seconds after the onset of the audio beep.

We conducted a pilot study to see if there were ceiling effects in the baseline condition, as well as to balance the difficulty level to ensure consistency before running the study to test our TOR design. This piloting allowed us to see issues early on. If a ceiling effect was detected, the scenario's critical situation was made more difficult to increase the range of possible responses of the participants. As a result, some problems occur in more scenarios then others, but there is no single problem that occurs in more than 6 scenarios.
\begin{table}
  \caption{Elements used to create scenarios}
  \label{table:1}
  \begin{tabular}{ccl}
  \toprule
  Technical problem & Weather & Obstacles  \\
  \midrule
  Low battery(6) & Fog(3), wind(4), snow(3) & Trees(2), buildings(3)\\
  Damaged propeller(5) & above water/sea(1) & Tower(1) \\
  Damaged camera(3) & High altitude(2) & Mountain(4) \\
  \bottomrule
  \end{tabular}
\end{table}

\begin{figure}[h]
  \centering
  \includegraphics[width=5in]{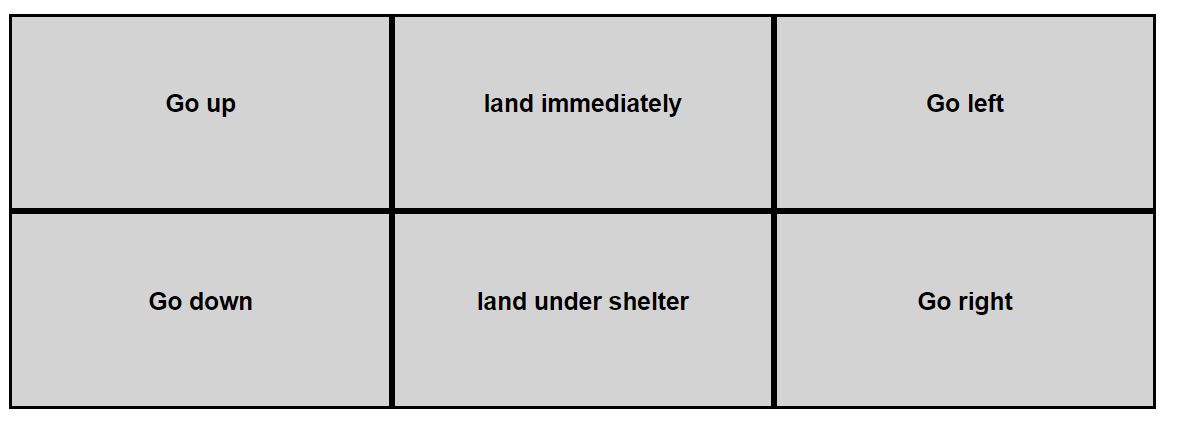}
  \caption{Possible options to resolve the critical situation}
  \Description{}
  \label{fig:Options}
\end{figure}

The goal of this user study was to assess whether participants would be able to use the critical information in order to arrive at a viable solution for the situation. To be able to test this, we presented 6 solution options (Figure \ref{fig:Options}) from which participants were required to choose. This measure is a proxy for performance, compared to giving participants access to the drone's remote control. It has the advantage that participants do not have to be experienced in drone flying to take part in the study, and it concentrates on the decision-making and not on other performance aspects (e.g. is the participant able to properly land the drone).
In a given scenario, not all of the presented options were suitable; there was always at least one correct and one incorrect option. While the options referring to changes in direction (up, down, right, left) stayed constant across scenarios, the two landing options varied between scenarios with at least one specific landing location given (e.g. land under the bridge). These landing locations did not have to be suitable landing locations. 

\subsection{Participants}
The study was hosted online on Lingoturk, which is an open-source crowd-sourcing client/server system that functioned as an interface to Prolific Academic. Prolific was used to recruit participants based on our prerequisites. Prerequisites to participate in the study were to be at least 18 years old and a native English speaker. Participants were not expected to have experience with drone piloting.
A total of 400 participants took part in this experiment. Every condition contained two lists, each with ten scenarios, and 40 participants took part in each list. They had not taken part in any previous relevant studies, and because we used a between-subject design, each participant only took part in one of the conditions. The experiment lasted about 14 minutes and was compensated with \textsterling 1.75.

\subsection{Experimental procedure} 
The participants were in the role of the controller and were allowed to decide how to resolve the critical situations. 
After giving informed consent to take part in the study, participants received detailed instructions and watched a demo video which was similar to the experimental videos detailed in the following. Participants here did not have to supply their own response, instead, the correct solution along with an explanation was presented. During the main part of the experiment, participants watched a series of simulated drone videos of scenarios in which the drone got into a difficult situation including a TOR at the end of the video. After each video, the participant was asked to take over control, by selecting one out of six options to proceed (Figure \ref{fig:Options}) indicated by a mouse click on the intended button. They were made aware of their decision being time sensitive by a counter displayed in the upper right corner that counted down from 7 seconds. If a participant failed to make a decision during this pre-specified take-over time, the screen timed out and the participant was asked for the reason (not enough time, none of the solutions fits, technical difficulties, or other).
After they made their decision, we asked two 5-point Likert scale questions to assess the participant's user experience: (1) Were you able to recognize the critical situation? (2) How obvious was the solution?. After finishing their set of experimental items, participants were automatically redirected to prolific to receive their payment.

\begin{figure}[h]
  \centering
  \includegraphics[width=3in]{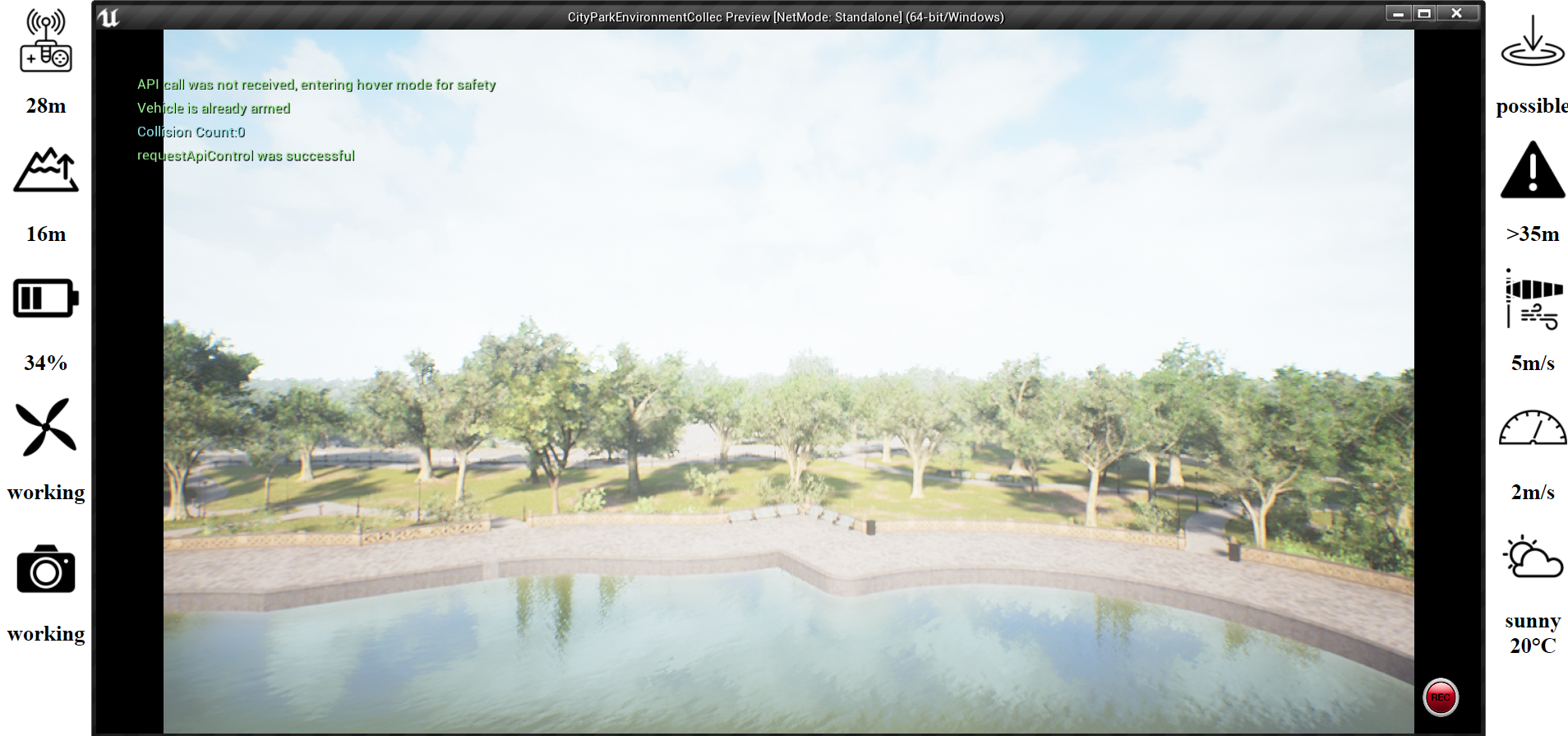}
  \caption{Screenshot of user interface}
  \Description{}
  \label{fig:screenshot}
\end{figure}

\subsection{Instructions to Participants}\label{sec:instructions}
The participant was informed that they are in the role of the controller of a semi-autonomous drone and have to decide how to resolve a critical situation. As participants were untrained and were expected to have little or no prior experience with drones, instructions were designed to enable the participant to make the correct decision in each scenario.

First, they were familiarized with the icons and their meaning and the general procedure of the experiment (i.e. drone flying autonomously, encountering a critical situation, take-over request, take-over by choosing suitable option). When explaining the different solution options, it was pointed out that landing immediately should only be chosen if this was possible (this information was provided in the respective icon) and that at least one specific landing location is always suggested, even if none are suitable for landing. We then informed the participant about what kinds of critical situations to expect and how to deal with them. According to their instructions, rain and snow would lead to internal damage as the drone is not waterproof. Physical damage was warned against when encountering obstacles or flying close to the ground. It was mentioned that changing the altitude could mitigate problems with fog (thicker close to the ground) and wind (stronger at high altitudes). Finally, it was pointed out that if either a propeller or camera is damaged, the drone has less precision in flying and that therefore obstacles might be more problematic than usual and landing manoeuvres might require more open space. These pieces of information were expected to be combined to find a correct solution for each scenario. Finally, participants were made aware of the need to respond within the 7-second time window after the video was completed.

\subsection{Experimental design} 
The experiment tested, in a between-participant design, the 5 different TOR designs presented in Section \ref{sec:TORdesign}.

\subsubsection{Variables and Hypothesis}
To be able to address Hypotheses H1--H3 as presented in Section \ref{sec:intro}, we use
the type of TOR as the independent variable. The dependent variables are (1) the decision accuracy, (2) response time, and (3) user experience as measured in the two Likert scale questions. Since we have 3 dependent measures, we formulate more specific sub-hypotheses for each hypothesis as follows:

Our main hypothesis H1 is that language+visual information TORs will improve the take-over process, as measured by the three dependent variables:
\newline \textbf{H1a}: Decision accuracy will be increased using TOR 3 (language + visual information) compared to TOR 1 (baseline) and TOR 2 (visual only).
\newline \textbf{H1b}: Response time will be shorter in TOR 3 compared to TOR 1 and TOR 2.
\newline \textbf{H1c}: Self-assessment of having recognized the critical situation and of the solution being obvious will be increased in TOR 3 compared to TOR 1 and TOR 2.

The first secondary hypothesis is that using fragments instead of full sentences will improve the take-over.
\newline \textbf{H2a}: Decision accuracy will be increased in TOR 4 (language (fragments)) compared to TOR 3 (language (sentences)).
\newline \textbf{H2b}: Response time will be shorter in TOR 4 compared to TOR 3.
\newline \textbf{H2c}: Self-assessment of having recognized the critical situation and of the solution being obvious will be increased in TOR 4 compared to TOR 3.

Lastly, we hypothesize that synchronizing the visual with the auditory information will facilitate information processing by the participants and thus lead to an improved take-over process:
\newline \textbf{H3a}: Decision accuracy will be increased in TOR 5 (language + visual with synchronous highlighting) compared to TOR 3 (language + visual with asynchronous highlighting).
\newline \textbf{H3b}: Response time will be shorter in TOR 5 compared to TOR 3.
\newline \textbf{H3c}: Self-assessment of having recognized the critical situation and of the solution being obvious will be increased in TOR 5 compared to TOR 3.

\subsubsection{Evaluation metrics and data analysis}
All analyses were carried out in R. Logistic regression mixed-effects models used the blme package which implements a Bayesian estimation method \citep{blme}, to improve convergence for random slope models. Linear mixed effects models were fitted using the package lme4 \cite{lme4}; significance for regression coefficients is determined using the Satterthwaite approximation of degrees of freedom provided by the package lmertest.

\textbf{Decision accuracy}: 
Decision accuracy was coded as correct if the solution selected by the participant was one of the intended correct solutions for the scenario, and incorrect if not. In cases where the participant did not pick a solution within the take-over time, the participant was prompted for the reason (\textit{not enough time}, \textit{none of the solutions fit}, \textit{technical difficulties}, or \textit{other}). While \textit{technical difficulties} were coded as NA, all other options were coded as incorrect, as the user did not arrive at a correct solution before the culmination of the critical situation.

For the inferential analyses, we used logistic mixed effects models with accuracy (correct or incorrect) as the outcome and TOR type as the fixed effect factor with 4 levels. Scenarios and Participants were treated as random effects to account for dependencies between data from the same sampling unit. We also included random slopes for TOR type within scenario to account for the expectation that not all scenarios would be equally susceptible to the TOR manipulation. Including random slopes reduces the risk that a detected significant fixed effect of TOR is in fact dependent on individual scenarios showing a particularly large effect. As our scenarios are not a representative sample of real-world drone piloting situations, we do not interpret the magnitude of the effect, but concentrate on the significance.

We conducted three analyses to address the three hypotheses stated above, where the first analysis, testing H1a, compares TOR 3 (language (sentence) + visual with asynchronous highlighting) to TOR 1 (baseline) and to TOR 2 (visual only). In the next step, we test H2a by contrasting TOR 3 (language (sentence) + visual with asynchronous highlighting) to TOR 4 (language (fragments) + visual with asynchronous highlighting). Finally, H3a is tested by comparing TOR 3 (language (sentence) + visual with asynchronous highlighting) with TOR 5 (language (sentence) + visual with synchronous highlighting).

\textbf{Response time}: Response time was defined as time between the end of the take-over request (equivalent to the onset of the take-over time and the presentation of the options) and the actual decision, which is a button press. If the participants did not respond to a TOR, their response time was adjusted to 7 seconds to analyze the data. Participants were not instructed to react as fast as they could, but just within the 7-second time frame, which limits comparability between response times in our study and other studies.

For the inferential analyses, we used linear mixed effect models with response time as the outcome and condition as the fixed effect. Scenarios and participants were treated as random effects. 

\textbf{User Experience}: Participants were asked two 5-point Likert scale questions after each video/scenario to provide feedback on their experience. The questions were formulated as follows:
    \begin{enumerate}
    \item[Q1:] Did you recognize the critical situation? \\
    Scale: 1: yes absolutely, 2: tentatively yes, 3: not sure, 4: tentatively no, 5: definitely not
    \item[Q2:] How obvious was the solution?\\
    Scale: 1: very obvious, 2: obvious, 3: not sure, 4: not so obvious, 5: not obvious at all 
    \end{enumerate}
The answers were coded numerically.

For the inferential analyses, we used linear mixed effect models with the rating as the outcome and condition as the fixed effect for both questions individually, thus treating the outcome as a continuous measure. Scenarios and Participants were treated as random effects. 


\section{Results}

In this section, we will in turn report our results regarding the three questions raised in this paper: whether there is a significant difference in accuracy, reaction time or user preference between the modalities (baseline, visual only vs.~visual and language), whether there is a difference between full sentences vs.~fragments, and whether the information in the language and visual highlighting modalities should be synchronized such that the highlighting comes on when it is mentioned in the spoken message.
\subsection{Missing data}
For the first 3 conditions (TORS 1-3), participants did not make a decision before time-out in 113 trials, corresponding to below 5\% of all trials. Table \ref{table:missing_full} gives the distribution by condition and differentiating different causes of the non-response as given by the participant. Only non-responses associated with technical difficulties were excluded from the analysis, while the other two categories were coded as incorrect. The overall rate of time-outs attributed to the decision period being too short was only 3\%, which suggests that 7 seconds was in general sufficient to make the decision.

\begin{table}[H]
  \caption{Non-responses by condition and reason for non-response for 3 conditions}
  \label{table:missing_full}
  \begin{tabular}{lccc}
  \toprule
  Design of TOR & ``not enough time'' & ``technical difficulty'' & ''none of the solutions fit`` \\
  \midrule
  (1) Baseline & 19 (2\%) & 6 (1\%) & 9 (1\%)\\
  (2) Visual Highlighting Only & 29 (4\%) & 3 (<1\%) & 13 (2\%)\\
  (3) Language + Visual \\ (asynchronous highlighting) & 27 (3\%) & 1 (<1\%) & 6 (1\%)\\
  \bottomrule
  \end{tabular}
\end{table}

\subsection{Analysis on visual highlighting plus spoken utterance vs. no highlighting and visual highlighting only}
We first compared the results from the baseline condition, the visual highlighting only condition and the visual highlighting with language condition. Our hypothesis had been that visual highlighting should help users to notice the critical situation, and that a spoken utterance might draw user attention even more effectively. We indeed find that decision accuracy is lowest in the baseline condition (48.74\%), while it is slightly higher when visual highlighting is included (51.32\%) and highest with both modalities (57.32\%). When inspecting the results for a more qualitative analysis to better understand what information participants may miss in the baseline condition, and when highlighting and the spoken utterance is most effective, we noticed a substantial amount of variability between items. We will therefore proceed to discuss these qualitative differences first, before proceeding with further analyses.


\subsubsection{Qualitative Analysis}\label{sec:qual-analysis}
We found three different types of scenarios: 
\begin{enumerate}
\item Scenarios where decision accuracy was improved by drawing participants' attention to the critical information (13 scenarios).  
\item Scenarios where information leads to an incorrect decision (5 scenarios): In scenarios where the propeller or the camera was damaged and there was an obstacle in the drone's flight path, many participants chose to ``land'' even if landing was not possible at the current location, as indicated by the corresponding icon. Here, performance can be lower in the highlighting or language utterance conditions, as participants perceive the critical information, but react to it incorrectly. We argue that in these cases, the actual goal of drawing the user's attention to the criticality has been successfully reached, but our accuracy measure does not reflect it. It also draws attention to the importance of training participants: They may not have paid attention to the icon signaling the possibility of immediate landing, as it was not part of the critical information that led to the TOR and hence was neither highlighted nor verbalized.
\item Scenarios with unclear landing options (2 scenarios): Here, the participants did not correctly understand their landing options, or did not know how much battery would be needed in order to reach a specific landing spot. Similar to the previous case, participants change their behavior based on the highlighted information, but do not decide correctly, or as intended.
\end{enumerate}

The qualitative analysis validates the effectiveness of the visual and linguistic messages, but also highlights the fact that perceiving a piece of information and making a decision regarding how to deal with it are two separate processes. In our study, we were first and foremost interested in whether the different types of presentation options effectively draw attention to the factors that contributed to the criticality of the situation. Therefore, we conducted subsequent analyses only on the subset of items for which noticing the information and deciding correctly fell together. 
In particular, we excluded all scenarios in which the camera or the propeller was broken and there was an obstacle in the drone's flight path, while landing was impossible (5 items). We also excluded scenarios where the landing location or the wording of the landing spot was unclear (2 items).

\subsubsection{Subset Analysis}
After excluding problematic scenarios, we analyzed the remaining set of 13 scenarios.

\begin{figure}
    \centering
    \includegraphics[scale=.6]{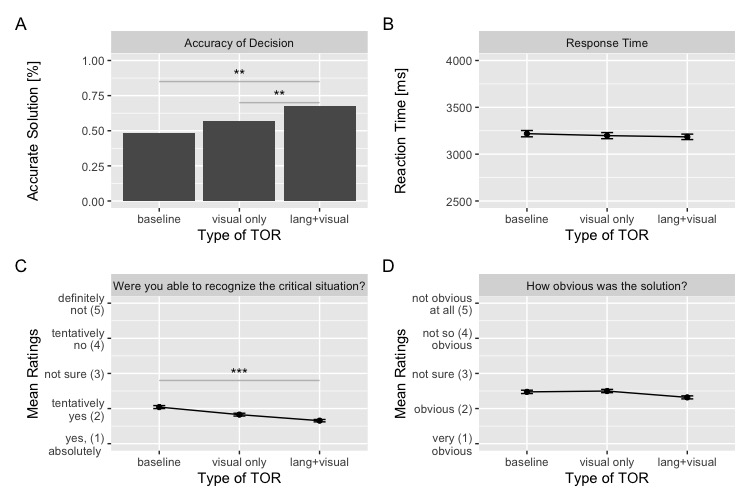}
    \caption{Summary graphs for analysis subset of 13 scenarios. A: Percentage of accurate solutions selected by participants by condition. B: Mean response times by condition. C: Participants' mean rating of 5-point Likert scale question on recognizability of criticality. D: Participants' mean rating of 5-point Likert scale question on obviousness of solution. Error bars indicate the standard error of the mean. Contrasts that showed significance in linear mixed models are indicated by stars.}
    \label{fig:subset_123}
\end{figure}

\paragraph{Decision accuracy}
The decision accuracy results for the remaining set of 13 scenarios are consistent with the accuracy results on the full dataset, but show a larger gap between performance in the baseline condition (48.74\%), the visual highlighting condition (56.86\%) and the condition with both modalities (67.63\%) (see also Figure \ref{fig:subset_123} panel A).

A logistic regression mixed effects model with random intercepts and slopes for condition under item, and random intercepts under subjects, confirms that the differences in decision accuracy between the language plus visual highlighting condition are significant compared to the baseline (z=2.950, p$<$0.01), as well as compared to the visual only condition (z=2.706, p$<$0.01). We can conclude that our hypothesis H1a is confirmed: Decision accuracy is increased if the TOR uses visual highlighting and a spoken message.

\paragraph{Response time}
The mean response times for the three TOR conditions are shown in Figure \ref{fig:subset_123}, panel B. The mean response time is around 3s in all the conditions. The linear mixed effect model did not reveal a statistically significant difference between the language plus visual highlighting and the other two conditions (t<1). As a result, hypothesis H1b (response times are shorter with language and visual highlighting) is rejected.

\paragraph{Likert Scale Questions}
Mean ratings of the Likert scale questions across the three conditions are given in Figure \ref{fig:subset_123} panels C and D. For question 1, asking for participants' self-assessment of having recognized the critical situation (1=yes, absolutely, 5=definitely not), our linear mixed effects analysis showed that participants felt they were more likely to have recognized the critical situation in the language plus visual highlighting condition compared to the baseline (t(236)=4.175, p$<$.001). The difference in language plus visual and visual only did not reach significance (t(236)=-1.679, p=.095).  







Question 2 asked participants how obvious they thought the solution was. Here, we find that scores are numerically better in the language plus visual highlighting condition than in the baseline and visual highlighting only conditions, but these differences did not reach statistical significance in our linear mixed effects model analysis (t(235)=1.597, p=0.112; t(237)=1.766, p=0.079). 
Hypothesis H1c is not confirmed, as we only see a difference in recognizing the critical situation, not in the obviousness of the solution.





\subsection{Analysis on fragments vs. sentences}
Our second question concerned the use of full sentences vs. shorter fragments for the spoken utterances. In this analysis, we compare the results for these types of formulations for the subset of 13 scenarios. 

\subsubsection{Decision accuracy}
As shown in Figure \ref{fig:subsetFS} Panel A, accuracy was on average higher in the sentence condition than in the fragments condition. A linear mixed effects model showed that this difference was not statistically significant (z=-1.825, p$=$0.068). We can thus reject hypothesis H2a and observe that, numerically, sentences are in fact more beneficial with respect to decision accuracy,  compared to fragments.

\subsubsection{Response time}
\begin{figure}
    \centering
    \includegraphics[scale=0.55]{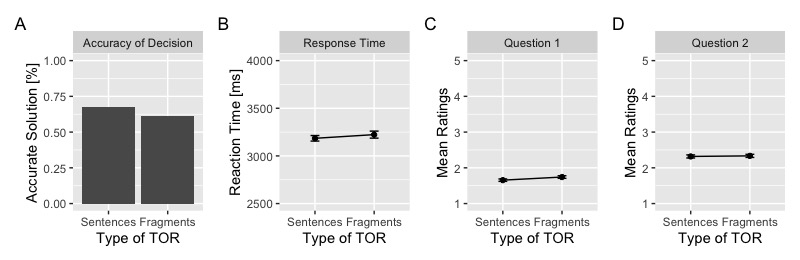}
    \caption{Summary graphs for analysis subset of 13 scenarios, comparing fragments to sentences. A: Percentage of accurate solutions selected by participants by condition. B: Mean response times by condition. C: Participants' mean rating of 5-point Likert scale question on recognizability of criticality. D: Participants' mean rating of 5-point Likert scale question on obviousness of solution. Low numbers indicate, respectively, that people felt they recognized the critical situation and that the solution was obvious. Error bars indicate the standard error of the mean.}
    \label{fig:subsetFS}
\end{figure}

Our second hypothesis concerned the response times -- we hypothesized that fragments might lead to faster decisions than full sentences, as they take less time to utter. 
The mean response times for the two conditions is shown in Panel B of Figure \ref{fig:subsetFS} -- it shows that the response time is very similar between the conditions; a linear mixed effects model shows no significant difference between conditions (t<1). We therefore conclude that fragments do not lead to faster responses, and we thus also reject H2b.








\subsubsection{Likert scale questions}
The two 5-point Likert scale questions that participants answered after each scenario concerning recognition of critical situation and obviousness of solution were investigated using linear mixed effect models. The analysis showed no significant effect of sentence vs.~fragment (t<1) for either of the questions. Numerically, sentences received slightly better ratings: see Panels C and D in Figure ~\ref{fig:subsetFS}. We can thus reject H2c.

Two additional questions on how comprehensible the message was and whether the length of the message was appropriate showed only minor differences, suggesting that both forms were perceived as appropriate formulations by the participants.

\subsection{Analysis on asynchronous vs. synchronous highlighting of icons}
Our third question was about synchronising the highlighting of visual icons with the spoken message. Due to an error in the design, no data was collected for one of the scenarios; we therefore compare the results of synchronous and asynchronous highlighting for a set of 12 scenarios. In synchronous highlighting, after the beep, a spoken message explains the take-over and the visual icons are highlighted when they are talked about and remain highlighted until the end of the TOR. Asynchronous highlighting, on the other hand, highlights the visual icons for the full duration of the TOR.

\subsubsection{Decision accuracy}
Figure \ref{fig:Synch} Panel A shows higher decision accuracy for the asynchronous highlighting condition compared to the synchronous highlighting condition. However, a linear mixed effects model shows no significant difference between conditions (z=-1.469, p=0.142). We can thus reject H3a with the additional note that numerically the effect is in the opposite direction.

\subsubsection{Response time}
\begin{figure}
    \centering
    \includegraphics[scale=0.55]{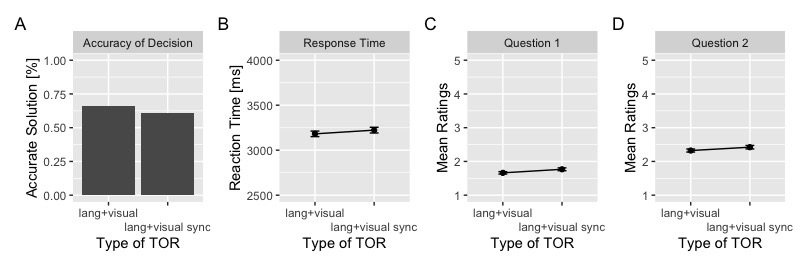}
    \caption{Summary graphs for analysis subset of 12 scenarios, comparing synchronized to asynchronous highlighting. A: Percentage of accurate solutions selected by participants by condition. B: Mean response times by condition. C: Participants' mean rating of 5-point Likert scale question on recognizability of criticality. D: Participants' mean rating of 5-point Likert scale question on obviousness of solution. Low numbers indicate, respectively, that people felt they recognized the critical situation and that the solution was obvious. Error bars indicate the standard error of the mean.}
    \label{fig:Synch}
\end{figure}

We hypothesised that synchronized highlighting would result in faster decisions than synchronized highlighting because the information is easier to process as the message is synchronized with highlighted icons. The mean response times for the two conditions is shown in Panel B of Figure \ref{fig:Synch}, which shows that synchronized highlighting leads to longer response time than asynchronous highlighting. A linear mixed effects model shows no significant difference between conditions (t(106)=1.88, p=0.063). As a result, we conclude that synchronization of the utterance with the visual highlighting of icons does not result in faster responses and we hence reject H3b.

\subsubsection{Likert scale questions}
The linear mixed effect models analysis shows that there is no statistically significant difference between conditions for both Question 1 (t(112)=1.032, p=0.304) and Question 2 (t<1). As a result, we can conclude that the timing of highlighting has no effect on the recognition of the critical situation and on the obviousness of the solution, and reject H3c. Numerically, we observe in Figure ~\ref{fig:Synch} that in comparison to the synchronous highlighting, participants in lang+visual (Language + Visual with asynchronous highlighting) gave better ratings with regard to recognizability and obviousness of the solution.

\section{Discussion, Limitations, Future Work}

Previous research on autonomous vehicles has found that multimodal take-over requests are more effective and that providing information to increase situational awareness can be beneficial. We designed a take-over request that takes these findings into account and extend them to a non-driving application. In particular, we used a drone setting as our testbed where the naive controller's ability to handle a critical situation depends heavily on their situational awareness. The take-over request consisted of an acoustic signal to get the controller's attention, which was followed by visual highlighting of sensor information as well as a spoken message verbalizing the same pieces of information. In the user study, this design was found to be beneficial for the take-over success, as evidenced by an increase in accuracy compared to the baseline take-over request which only consisted of an abstract acoustic signal, as well as compared to a condition where an abstract acoustic signal was coupled with information conveyed in the visual domain alone. At least one reason for this benefit can be found by looking at the measures of user experience, as participants reported on average better recognition of the critical situation in the target TOR than in the other conditions. 

Shortening the spoken message to fragments instead of full sentences did not result in increased accuracy, nor in faster responses, in our user study. While the difference was not significant, the numerical effect even went in the other direction: Sentences led to higher accuracy, suggesting that more condensed messages might be harder to process in this setting. 

Contrary to our expectations, synchronizing visual highlighting with the spoken message also did not result in an improvement, but in a numerical decrease in accuracy and a numerical increase in reaction time. Clearly, synchronization did not facilitate the processing of information. A possible cause could be that in the synchronized version the visual information was presented for a shorter time in total. Also, although synchronized with the spoken message, the timing of highlighting might have seemed counter-intuitive to some participants, thus causing more confusion than understanding. In particular, participants might expect to be given access to information as soon as it is available, as is the case e.g. with warning lights for water level, etc., in cars.
Finally, forcing the user to take up the information in a pre-specified order might interfere with the user's own goals and a fast establishment of situational awareness.

The qualitative analysis revealed that some participants did perceive the critical information, but then reacted incorrectly in specific cases of a broken camera and a broken propeller. In these cases, the participants predominantly decided to land immediately, even though landing was not possible in some of the affected scenarios, as indicated by one of the icons. For future work, an interesting direction would therefore be to not only consider   highlighting those icons that are relevant to the detected critical situation, but also icons that are particularly relevant to the next action options. 
Furthermore, we would recommend that for studies like ours, where the focus is most strongly on whether a piece of information is communicated successfully, it could be useful to tease apart the effect of alerting the user about a specific piece of information from the overall effect on decision accuracy. 

Response times, in our study, were on average around 3 seconds, with little variation between conditions. In contrast to other studies, participants were not instructed to respond as fast as they could, but were only made aware of the time limit of 7 seconds. What we found is thus that participants were able to respond slightly faster than expected, which provides more flexibility in the timing of the take-over process: Shortening the take-over period to 5 seconds seems a viable option.

Unfortunately, we were not able to conduct the study onsite because of restrictions caused by the COVID-19 pandemic. The results of this online study should be validated by a lab study that should feature a longer training phase and in which participants could select the preferred option on a separate screen to avoid the sudden cut-off of the video transmission. 
Additionally, the study could be run with an additional secondary task, to make sure that participants are immersed in another task and have less time to watch the drone and the icons over the course of the flight, such that initial situation awareness at the handover request signal would be even lower than in the present study.

At the end of our study, we asked participants for feedback on the study. A number of participants mentioned that it was difficult to choose the correct solution and that they would have appreciated if the message could give a hint toward this. This could be an interesting future direction in the design of TORs. For instance, if there are several alternative ways to handle a situation, and the semi-autonomous system is not in a position to decide between them, the system could mention them along with their advantages and problems.

While this study was conducted in the context of semi-autonomous drones, similar multi-modal interactions could be applied and evaluated in the domain of human-robot cooperation \cite{bannat2009multimodal,huber2009,moon2014,Strabala2013} and take-over procedures for semi-autonomous driving \cite{Borojeni2016,Jeon2019,McKeown,Politis2015,WiehrICMI2021}.
This is especially relevant in situations where the information that needs to be conveyed to the operator to achieve situation awareness is complex and where naive users are required to interact with cyber-physical systems.

\section{Conclusion}
In this study, we proposed a design of a take-over request in the setting of semi-autonomous drones and tested specific features relevant for take-over requests in general. The take-over request consists of an abstract auditory warning followed by visual highlighting of relevant sensor information together with a spoken message. After the request, sufficient time is given to the controller to process the information, before an action is required. This take-over design was tested in an online user study with naive participants that confirmed the effectiveness as evidenced by an increase of accuracy of the decision compared to simpler TOR designs which did not include the spoken message. Shortening of the message as well as an improved synchronization between the visual and the auditory modality in conveying information did not result in better accuracy, response times or better ratings in user's preferences.



\bibliographystyle{ACM-Reference-Format}
\bibliography{references-1,references-2}


\begin{thebibliography}{49}


\ifx \showCODEN    \undefined \def \showCODEN     #1{\unskip}     \fi
\ifx \showDOI      \undefined \def \showDOI       #1{#1}\fi
\ifx \showISBNx    \undefined \def \showISBNx     #1{\unskip}     \fi
\ifx \showISBNxiii \undefined \def \showISBNxiii  #1{\unskip}     \fi
\ifx \showISSN     \undefined \def \showISSN      #1{\unskip}     \fi
\ifx \showLCCN     \undefined \def \showLCCN      #1{\unskip}     \fi
\ifx \shownote     \undefined \def \shownote      #1{#1}          \fi
\ifx \showarticletitle \undefined \def \showarticletitle #1{#1}   \fi
\ifx \showURL      \undefined \def \showURL       {\relax}        \fi
\providecommand\bibfield[2]{#2}
\providecommand\bibinfo[2]{#2}
\providecommand\natexlab[1]{#1}
\providecommand\showeprint[2][]{arXiv:#2}

\bibitem[Adams et~al\mbox{.}(1995)]%
        {E2:Adams1995}
\bibfield{author}{\bibinfo{person}{Marilyn~Jager Adams},
  \bibinfo{person}{Yvette~J. Tenney}, {and} \bibinfo{person}{Richard~W. Pew}.}
  \bibinfo{year}{1995}\natexlab{}.
\newblock \showarticletitle{{Situation Awareness and the Cognitive Management
  of Complex Systems}}.
\newblock \bibinfo{journal}{\emph{Human Factors}} \bibinfo{volume}{37},
  \bibinfo{number}{1} (\bibinfo{year}{1995}), \bibinfo{pages}{85--104}.
\newblock
\showISBNx{0018-7208}
\showISSN{00187208}
\urldef\tempurl%
\url{https://doi.org/10.1518/001872095779049462}
\showDOI{\tempurl}


\bibitem[Ahirwar et~al\mbox{.}(2019)]%
        {Ahirwar}
\bibfield{author}{\bibinfo{person}{S Ahirwar}, \bibinfo{person}{Swarnkar
  Swarnkar}, \bibinfo{person}{Bhukya Srinivas}, {and} \bibinfo{person}{G.
  Namwade}.} \bibinfo{year}{2019}\natexlab{}.
\newblock \showarticletitle{Application of Drone in Agriculture}.
\newblock \bibinfo{journal}{\emph{International Journal of Current Microbiology
  and Applied Sciences}}  \bibinfo{volume}{8} (\bibinfo{date}{01}
  \bibinfo{year}{2019}), \bibinfo{pages}{2500--2505}.
\newblock
\urldef\tempurl%
\url{https://doi.org/10.20546/ijcmas.2019.801.264}
\showDOI{\tempurl}


\bibitem[Allopenna et~al\mbox{.}(1998)]%
        {Allopenna98}
\bibfield{author}{\bibinfo{person}{Paul~D. Allopenna},
  \bibinfo{person}{James~S. Magnuson}, {and} \bibinfo{person}{Michael~K.
  Tanenhaus}.} \bibinfo{year}{1998}\natexlab{}.
\newblock \showarticletitle{Tracking the Time Course of Spoken Word Recognition
  Using Eye Movements: Evidence for Continuous Mapping Models}.
\newblock \bibinfo{journal}{\emph{Journal of memory and language}}
  \bibinfo{volume}{38} (\bibinfo{year}{1998}), \bibinfo{pages}{419--439}.
\newblock


\bibitem[Bannat et~al\mbox{.}(2009)]%
        {bannat2009multimodal}
\bibfield{author}{\bibinfo{person}{Alexander Bannat},
  \bibinfo{person}{J{\"u}rgen Gast}, \bibinfo{person}{Tobias Rehrl},
  \bibinfo{person}{Wolfgang R{\"o}sel}, \bibinfo{person}{Gerhard Rigoll}, {and}
  \bibinfo{person}{Frank Wallhoff}.} \bibinfo{year}{2009}\natexlab{}.
\newblock \showarticletitle{A multimodal human-robot-interaction scenario:
  Working together with an industrial robot}. In
  \bibinfo{booktitle}{\emph{International conference on human-computer
  interaction}}. Springer, \bibinfo{pages}{303--311}.
\newblock


\bibitem[Bates et~al\mbox{.}(2015)]%
        {lme4}
\bibfield{author}{\bibinfo{person}{Douglas Bates}, \bibinfo{person}{Martin
  M{\"a}chler}, \bibinfo{person}{Ben Bolker}, {and} \bibinfo{person}{Steve
  Walker}.} \bibinfo{year}{2015}\natexlab{}.
\newblock \showarticletitle{Fitting Linear Mixed-effects Models Using {lme4}}.
\newblock \bibinfo{journal}{\emph{Journal of Statistical Software}}
  \bibinfo{volume}{67}, \bibinfo{number}{1} (\bibinfo{year}{2015}),
  \bibinfo{pages}{1--48}.
\newblock
\urldef\tempurl%
\url{https://doi.org/10.18637/jss.v067.i01}
\showDOI{\tempurl}


\bibitem[Bliss(2003)]%
        {Bliss2003}
\bibfield{author}{\bibinfo{person}{James~P. Bliss}.}
  \bibinfo{year}{2003}\natexlab{}.
\newblock \showarticletitle{Investigation of Alarm-related Accidents and
  Incidents in Aviation}.
\newblock \bibinfo{journal}{\emph{The International Journal of Aviation
  Psychology}} \bibinfo{volume}{13}, \bibinfo{number}{3}
  (\bibinfo{year}{2003}), \bibinfo{pages}{249--268}.
\newblock
\urldef\tempurl%
\url{https://doi.org/10.1207/S15327108IJAP1303_04}
\showDOI{\tempurl}


\bibitem[Bliss et~al\mbox{.}(1995)]%
        {Bliss1995}
\bibfield{author}{\bibinfo{person}{James~P. Bliss}, \bibinfo{person}{Richard~D.
  Gilson}, {and} \bibinfo{person}{John~E. Deaton}.}
  \bibinfo{year}{1995}\natexlab{}.
\newblock \showarticletitle{Human Probability Matching Behaviour in Response to
  Alarms of Varying Reliability}.
\newblock \bibinfo{journal}{\emph{Ergonomics}} \bibinfo{volume}{38},
  \bibinfo{number}{11} (\bibinfo{year}{1995}), \bibinfo{pages}{2300--2312}.
\newblock
\urldef\tempurl%
\url{https://doi.org/10.1080/00140139508925269}
\showDOI{\tempurl}


\bibitem[Borojeni et~al\mbox{.}(2016)]%
        {Borojeni2016}
\bibfield{author}{\bibinfo{person}{Shadan~Sadeghian Borojeni},
  \bibinfo{person}{Lewis Chuang}, \bibinfo{person}{Wilko Heuten}, {and}
  \bibinfo{person}{Susanne Boll}.} \bibinfo{year}{2016}\natexlab{}.
\newblock \showarticletitle{Assisting Drivers with Ambient Take-over Requests
  in Highly Automated Driving}. In \bibinfo{booktitle}{\emph{Proceedings of the
  8th International Conference on Automotive User Interfaces and Interactive
  Vehicular Applications}} (Ann Arbor, MI, USA)
  \emph{(\bibinfo{series}{Automotive'UI 16})}. \bibinfo{publisher}{Association
  for Computing Machinery}, \bibinfo{address}{New York, NY, USA},
  \bibinfo{pages}{237–244}.
\newblock
\showISBNx{9781450345330}
\urldef\tempurl%
\url{https://doi.org/10.1145/3003715.3005409}
\showDOI{\tempurl}


\bibitem[Borojeni et~al\mbox{.}(2018)]%
        {Borojeni2018}
\bibfield{author}{\bibinfo{person}{Shadan~Sadeghian Borojeni},
  \bibinfo{person}{Lars Weber}, \bibinfo{person}{Wilko Heuten}, {and}
  \bibinfo{person}{Susanne Boll}.} \bibinfo{year}{2018}\natexlab{}.
\newblock \showarticletitle{From Reading to Driving: Priming Mobile Users for
  Take-over Situations in Highly Automated Driving}. In
  \bibinfo{booktitle}{\emph{Proceedings of the 20th International Conference on
  Human-Computer Interaction with Mobile Devices and Services}} (Barcelona,
  Spain) \emph{(\bibinfo{series}{MobileHCI '18})}.
  \bibinfo{publisher}{Association for Computing Machinery},
  \bibinfo{address}{New York, NY, USA}, Article \bibinfo{articleno}{14},
  \bibinfo{numpages}{12}~pages.
\newblock
\showISBNx{9781450358989}
\urldef\tempurl%
\url{https://doi.org/10.1145/3229434.3229464}
\showDOI{\tempurl}


\bibitem[Bravo and Leiras(2015)]%
        {Bravo2015}
\bibfield{author}{\bibinfo{person}{Raissa Zurli~Bittencourt Bravo} {and}
  \bibinfo{person}{A. Leiras}.} \bibinfo{year}{2015}\natexlab{}.
\newblock \showarticletitle{Literature Review of the Applications of Uavs in
  Humanitarian Relief}.
\newblock


\bibitem[{Chang} et~al\mbox{.}(2022)]%
        {ChangLREC2022}
\bibfield{author}{\bibinfo{person}{Ernie {Chang}}, \bibinfo{person}{Alisa
  {Kovtunova}}, \bibinfo{person}{Stefan {Borgwardt}}, \bibinfo{person}{Vera
  {Demberg}}, \bibinfo{person}{Kathryn {Chapman}}, {and}
  \bibinfo{person}{Hui{-}Syuan {Yeh}}.} \bibinfo{year}{2022}\natexlab{}.
\newblock \showarticletitle{Logic-guided Message Generation from Raw Real-time
  Sensor Data}. In \bibinfo{booktitle}{\emph{Proceedings of the 13th Language
  Resources and Evaluation Conference of the European Language Resources
  Association (LREC)}}.
\newblock
\newblock
\shownote{To appear.}.


\bibitem[Chung et~al\mbox{.}(2013)]%
        {blme}
\bibfield{author}{\bibinfo{person}{Yeojin Chung}, \bibinfo{person}{Sophia
  Rabe-Hesketh}, \bibinfo{person}{Vincent Dorie}, \bibinfo{person}{Andrew
  Gelman}, {and} \bibinfo{person}{Jingchen Liu}.}
  \bibinfo{year}{2013}\natexlab{}.
\newblock \showarticletitle{A Nondegenerate Penalized Likelihood Estimator for
  Variance Parameters in Multilevel Models}.
\newblock \bibinfo{journal}{\emph{Psychometrika}} \bibinfo{volume}{78},
  \bibinfo{number}{4} (\bibinfo{year}{2013}), \bibinfo{pages}{685--709}.
\newblock
\urldef\tempurl%
\url{https://doi.org/10.1007/s11336-013-9328-2}
\showDOI{\tempurl}


\bibitem[Coffey(2014)]%
        {Coffey}
\bibfield{author}{\bibinfo{person}{Valerie Coffey}.}
  \bibinfo{year}{2014}\natexlab{}.
\newblock \showarticletitle{High-energy Lasers: New Advances in Defense
  Applications}.
\newblock  (\bibinfo{year}{2014}).
\newblock


\bibitem[Cohen and McGee(2004)]%
        {E2:cohen2004tangible}
\bibfield{author}{\bibinfo{person}{Philip~R. Cohen} {and}
  \bibinfo{person}{David~R. McGee}.} \bibinfo{year}{2004}\natexlab{}.
\newblock \showarticletitle{Tangible Multimodal Interfaces for Safety-critical
  Applications}.
\newblock \bibinfo{journal}{\emph{Commun. ACM}} \bibinfo{volume}{47},
  \bibinfo{number}{1} (\bibinfo{year}{2004}), \bibinfo{pages}{41--46}.
\newblock


\bibitem[Endsley(1996)]%
        {E2:endsley1996automation}
\bibfield{author}{\bibinfo{person}{Mica~R. Endsley}.}
  \bibinfo{year}{1996}\natexlab{}.
\newblock \showarticletitle{Automation and Situation Awareness}.
\newblock \bibinfo{journal}{\emph{Automation and Human Performance: Theory and
  Applications}} (\bibinfo{year}{1996}), \bibinfo{pages}{163--181}.
\newblock


\bibitem[Flammini et~al\mbox{.}(2016)]%
        {Flammini}
\bibfield{author}{\bibinfo{person}{Francesco Flammini},
  \bibinfo{person}{Concetta Pragliola}, {and} \bibinfo{person}{Giovanni
  Smarra}.} \bibinfo{year}{2016}\natexlab{}.
\newblock \showarticletitle{Railway Infrastructure Monitoring by Drones}.
  \bibinfo{pages}{1--6}.
\newblock
\urldef\tempurl%
\url{https://doi.org/10.1109/ESARS-ITEC.2016.7841398}
\showDOI{\tempurl}


\bibitem[Gold et~al\mbox{.}(2013)]%
        {Gold2013}
\bibfield{author}{\bibinfo{person}{Christian Gold}, \bibinfo{person}{Daniel
  Damb{\"o}ck}, \bibinfo{person}{Lutz Lorenz}, {and} \bibinfo{person}{Klaus
  Bengler}.} \bibinfo{year}{2013}\natexlab{}.
\newblock \showarticletitle{“take Over!” How Long Does It Take to Get the
  Driver Back into the Loop?}. In \bibinfo{booktitle}{\emph{Proceedings of the
  human factors and ergonomics society annual meeting}},
  Vol.~\bibinfo{volume}{57}. Sage Publications Sage CA: Los Angeles, CA,
  \bibinfo{pages}{1938--1942}.
\newblock


\bibitem[Haeuslschmid et~al\mbox{.}(2017)]%
        {E2:pflegingetal17}
\bibfield{author}{\bibinfo{person}{Renate Haeuslschmid}, \bibinfo{person}{Max
  von Buelow}, \bibinfo{person}{Bastian Pfleging}, {and}
  \bibinfo{person}{Andreas Butz}.} \bibinfo{year}{2017}\natexlab{}.
\newblock \showarticletitle{Supporting Trust in Autonomous Driving}. In
  \bibinfo{booktitle}{\emph{International Conference on Intelligent User
  Interfaces (IUI)}}. \bibinfo{pages}{319--329}.
\newblock
\urldef\tempurl%
\url{https://doi.org/10.1145/3025171.3025198}
\showDOI{\tempurl}


\bibitem[Hancock(2017)]%
        {Hancock2017}
\bibfield{author}{\bibinfo{person}{Peter~A Hancock}.}
  \bibinfo{year}{2017}\natexlab{}.
\newblock \showarticletitle{Driven to Distraction and Back Again}.
\newblock In \bibinfo{booktitle}{\emph{Driver Distraction and Inattention}}.
  \bibinfo{publisher}{CRC Press}, \bibinfo{pages}{9--26}.
\newblock


\bibitem[Hassenzahl and Klapperich(2014)]%
        {Hassenzahl}
\bibfield{author}{\bibinfo{person}{Marc Hassenzahl} {and}
  \bibinfo{person}{Holger Klapperich}.} \bibinfo{year}{2014}\natexlab{}.
\newblock \showarticletitle{Convenient, Clean, and Efficient? The Experiential
  Costs of Everyday Automation}. In \bibinfo{booktitle}{\emph{Proceedings of
  the 8th Nordic Conference on Human-Computer Interaction: Fun, Fast,
  Foundational}} (Helsinki, Finland) \emph{(\bibinfo{series}{NordiCHI '14})}.
  \bibinfo{publisher}{Association for Computing Machinery},
  \bibinfo{address}{New York, NY, USA}, \bibinfo{pages}{21–30}.
\newblock
\showISBNx{9781450325424}
\urldef\tempurl%
\url{https://doi.org/10.1145/2639189.2639248}
\showDOI{\tempurl}


\bibitem[Huber et~al\mbox{.}(2008)]%
        {huber2009}
\bibfield{author}{\bibinfo{person}{Markus Huber}, \bibinfo{person}{Markus
  Rickert}, \bibinfo{person}{Alois Knoll}, \bibinfo{person}{Thomas Brandt},
  {and} \bibinfo{person}{Stefan Glasauer}.} \bibinfo{year}{2008}\natexlab{}.
\newblock \showarticletitle{Human-robot interaction in handing-over tasks}. In
  \bibinfo{booktitle}{\emph{RO-MAN 2008 - The 17th IEEE International Symposium
  on Robot and Human Interactive Communication}}. \bibinfo{pages}{107--112}.
\newblock
\urldef\tempurl%
\url{https://doi.org/10.1109/ROMAN.2008.4600651}
\showDOI{\tempurl}


\bibitem[Jeon(2019)]%
        {Jeon2019}
\bibfield{author}{\bibinfo{person}{Myounghoon Jeon}.}
  \bibinfo{year}{2019}\natexlab{}.
\newblock \showarticletitle{Multimodal Displays for Take-over in Level 3
  Automated Vehicles While Playing a Game}. In
  \bibinfo{booktitle}{\emph{Extended Abstracts of the 2019 CHI Conference on
  Human Factors in Computing Systems}} (Glasgow, Scotland Uk)
  \emph{(\bibinfo{series}{CHI EA '19})}. \bibinfo{publisher}{Association for
  Computing Machinery}, \bibinfo{address}{New York, NY, USA},
  \bibinfo{pages}{1–6}.
\newblock
\showISBNx{9781450359719}
\urldef\tempurl%
\url{https://doi.org/10.1145/3290607.3313056}
\showDOI{\tempurl}


\bibitem[Kaleem et~al\mbox{.}(2018)]%
        {Kaleem}
\bibfield{author}{\bibinfo{person}{Zeeshan Kaleem},
  \bibinfo{person}{Mubashir~Husain Rehmani}, \bibinfo{person}{Ejaz Ahmed},
  \bibinfo{person}{Abbas Jamalipour}, \bibinfo{person}{Joel~JPC Rodrigues},
  \bibinfo{person}{Hassna Moustafa}, {and} \bibinfo{person}{Wael Guibene}.}
  \bibinfo{year}{2018}\natexlab{}.
\newblock \showarticletitle{Amateur Drone Surveillance: Applications,
  Architectures, Enabling Technologies, and Public Safety Issues: Part 1}.
\newblock \bibinfo{journal}{\emph{IEEE Communications Magazine}}
  \bibinfo{volume}{56}, \bibinfo{number}{1} (\bibinfo{year}{2018}),
  \bibinfo{pages}{14--15}.
\newblock


\bibitem[Kalnikaite et~al\mbox{.}(2011)]%
        {E2:kalnikaite2011nudge}
\bibfield{author}{\bibinfo{person}{Vaiva Kalnikaite}, \bibinfo{person}{Yvonne
  Rogers}, \bibinfo{person}{Jon Bird}, \bibinfo{person}{Nicolas Villar},
  \bibinfo{person}{Khaled Bachour}, \bibinfo{person}{Stephen Payne},
  \bibinfo{person}{Peter~M. Todd}, \bibinfo{person}{Johannes Sch{\"o}ning},
  \bibinfo{person}{Antonio Kr{\"u}ger}, {and} \bibinfo{person}{Stefan
  Kreitmayer}.} \bibinfo{year}{2011}\natexlab{}.
\newblock \showarticletitle{How to Nudge in Situ: {D}esigning Lambent Devices
  to Deliver Salient Information in Supermarkets}. In
  \bibinfo{booktitle}{\emph{International Conference on Ubiquitous Computing
  (UbiComp)}}. \bibinfo{pages}{11--20}.
\newblock


\bibitem[Khosravi et~al\mbox{.}(2021)]%
        {Khosravi}
\bibfield{author}{\bibinfo{person}{Mohammadjavad Khosravi},
  \bibinfo{person}{Saeede Enayati}, \bibinfo{person}{Hamid Saeedi}, {and}
  \bibinfo{person}{Hossein Pishro-Nik}.} \bibinfo{year}{2021}\natexlab{}.
\newblock \showarticletitle{Multi-purpose Drones for Coverage and Transport
  Applications}.
\newblock \bibinfo{journal}{\emph{IEEE Transactions on Wireless
  Communications}} (\bibinfo{year}{2021}), \bibinfo{pages}{1--1}.
\newblock
\urldef\tempurl%
\url{https://doi.org/10.1109/TWC.2021.3054748}
\showDOI{\tempurl}


\bibitem[Koo et~al\mbox{.}(2014)]%
        {Koo}
\bibfield{author}{\bibinfo{person}{Jeamin Koo}, \bibinfo{person}{Jungsuk Kwac},
  \bibinfo{person}{Martin Steinert}, \bibinfo{person}{Larry Leifer}, {and}
  \bibinfo{person}{Clifford Nass}.} \bibinfo{year}{2014}\natexlab{}.
\newblock \showarticletitle{Why Did My Car Just Do That? Explaining
  Semi-autonomous Driving Actions to Improve Driver Understanding, Trust, and
  Performance}.
\newblock \bibinfo{journal}{\emph{International Journal on Interactive Design
  and Manufacturing (IJIDeM)}}  \bibinfo{volume}{9} (\bibinfo{date}{01}
  \bibinfo{year}{2014}).
\newblock
\urldef\tempurl%
\url{https://doi.org/10.1007/s12008-014-0227-2}
\showDOI{\tempurl}


\bibitem[K{\"o}rber et~al\mbox{.}(2018)]%
        {Krber2018}
\bibfield{author}{\bibinfo{person}{Moritz K{\"o}rber}, \bibinfo{person}{Lorenz
  Prasch}, {and} \bibinfo{person}{Klaus Bengler}.}
  \bibinfo{year}{2018}\natexlab{}.
\newblock \showarticletitle{Why Do I Have to Drive Now? Post Hoc Explanations
  of Takeover Requests}.
\newblock \bibinfo{journal}{\emph{Human Factors The Journal of the Human
  Factors and Ergonomics Society}}  \bibinfo{volume}{60} (\bibinfo{date}{05}
  \bibinfo{year}{2018}), \bibinfo{pages}{305–323}.
\newblock
\urldef\tempurl%
\url{https://doi.org/10.1177/0018720817747730}
\showDOI{\tempurl}


\bibitem[Maghazei and Netland(2019)]%
        {Maghazei}
\bibfield{author}{\bibinfo{person}{Omid Maghazei} {and}
  \bibinfo{person}{Torbj{\o}rn~H. Netland}.} \bibinfo{year}{2019}\natexlab{}.
\newblock \showarticletitle{Drones in Manufacturing: Exploring Opportunities
  for Research and Practice}.
\newblock \bibinfo{journal}{\emph{Journal of Manufacturing Technology
  Management}} (\bibinfo{year}{2019}).
\newblock


\bibitem[McKeown and Isherwood(2007)]%
        {McKeown}
\bibfield{author}{\bibinfo{person}{Denis McKeown} {and} \bibinfo{person}{Sarah
  Isherwood}.} \bibinfo{year}{2007}\natexlab{}.
\newblock \showarticletitle{Mapping Candidate Within-vehicle Auditory Displays
  to Their Referents}.
\newblock \bibinfo{journal}{\emph{Human Factors}} (\bibinfo{year}{2007}),
  \bibinfo{pages}{417--428}.
\newblock


\bibitem[Moon et~al\mbox{.}(2014)]%
        {moon2014}
\bibfield{author}{\bibinfo{person}{AJung Moon}, \bibinfo{person}{Daniel~M.
  Troniak}, \bibinfo{person}{Brian Gleeson}, \bibinfo{person}{Matthew~K.X.J.
  Pan}, \bibinfo{person}{Minhua Zheng}, \bibinfo{person}{Benjamin~A. Blumer},
  \bibinfo{person}{Karon MacLean}, {and} \bibinfo{person}{Elizabeth~A. Croft}.}
  \bibinfo{year}{2014}\natexlab{}.
\newblock \showarticletitle{Meet Me Where i'm Gazing: How Shared Attention Gaze
  Affects Human-Robot Handover Timing}. In
  \bibinfo{booktitle}{\emph{Proceedings of the 2014 ACM/IEEE International
  Conference on Human-Robot Interaction}} (Bielefeld, Germany)
  \emph{(\bibinfo{series}{HRI '14})}. \bibinfo{publisher}{Association for
  Computing Machinery}, \bibinfo{address}{New York, NY, USA},
  \bibinfo{pages}{334–341}.
\newblock
\showISBNx{9781450326582}
\urldef\tempurl%
\url{https://doi.org/10.1145/2559636.2559656}
\showDOI{\tempurl}


\bibitem[M{\"u}ller et~al\mbox{.}(2009)]%
        {E2:muller2009reflectivesigns}
\bibfield{author}{\bibinfo{person}{J{\"o}rg M{\"u}ller},
  \bibinfo{person}{Juliane Exeler}, \bibinfo{person}{Markus Buzeck}, {and}
  \bibinfo{person}{Antonio Kr{\"u}ger}.} \bibinfo{year}{2009}\natexlab{}.
\newblock \showarticletitle{Reflectivesigns: Digital Signs That Adapt to
  Audience Attention}. In \bibinfo{booktitle}{\emph{IEEE International
  Conference on Pervasive Computing and Communications (PerCom)}}.
  \bibinfo{pages}{17--24}.
\newblock
\urldef\tempurl%
\url{https://doi.org/10.1007/978-3-642-01516-8_3}
\showDOI{\tempurl}


\bibitem[Naujoks et~al\mbox{.}(2014)]%
        {Naujoks2014}
\bibfield{author}{\bibinfo{person}{Frederik Naujoks},
  \bibinfo{person}{Christoph Mai}, {and} \bibinfo{person}{A. Neukum}.}
  \bibinfo{year}{2014}\natexlab{}.
\newblock \showarticletitle{The Effect of Urgency of Take-over Requests during
  Highly Automated Driving under Distraction Conditions}.
\newblock


\bibitem[Politis et~al\mbox{.}(2013)]%
        {Politis2013}
\bibfield{author}{\bibinfo{person}{Ioannis Politis}, \bibinfo{person}{Stephen
  Brewster}, {and} \bibinfo{person}{Frank Pollick}.}
  \bibinfo{year}{2013}\natexlab{}.
\newblock \showarticletitle{Evaluating Multimodal Driver Displays of Varying
  Urgency}. In \bibinfo{booktitle}{\emph{Proceedings of the 5th International
  Conference on Automotive User Interfaces and Interactive Vehicular
  Applications}} (Eindhoven, Netherlands) \emph{(\bibinfo{series}{AutomotiveUI
  '13})}. \bibinfo{publisher}{Association for Computing Machinery},
  \bibinfo{address}{New York, NY, USA}, \bibinfo{pages}{92–99}.
\newblock
\showISBNx{9781450324786}
\urldef\tempurl%
\url{https://doi.org/10.1145/2516540.2516543}
\showDOI{\tempurl}


\bibitem[Politis et~al\mbox{.}(2015a)]%
        {Politis2015}
\bibfield{author}{\bibinfo{person}{Ioannis Politis}, \bibinfo{person}{Stephen
  Brewster}, {and} \bibinfo{person}{Frank Pollick}.}
  \bibinfo{year}{2015}\natexlab{a}.
\newblock \showarticletitle{Language-based Multimodal Displays for the Handover
  of Control in Autonomous Cars}. In \bibinfo{booktitle}{\emph{Proceedings of
  the 7th International Conference on Automotive User Interfaces and
  Interactive Vehicular Applications}} (Nottingham, United Kingdom)
  \emph{(\bibinfo{series}{AutomotiveUI '15})}. \bibinfo{publisher}{Association
  for Computing Machinery}, \bibinfo{address}{New York, NY, USA},
  \bibinfo{pages}{3–10}.
\newblock
\showISBNx{9781450337366}
\urldef\tempurl%
\url{https://doi.org/10.1145/2799250.2799262}
\showDOI{\tempurl}


\bibitem[Politis et~al\mbox{.}(2015b)]%
        {PolitisBeep}
\bibfield{author}{\bibinfo{person}{Ioannis Politis}, \bibinfo{person}{Stephen
  Brewster}, {and} \bibinfo{person}{Frank Pollick}.}
  \bibinfo{year}{2015}\natexlab{b}.
\newblock \showarticletitle{To Beep or Not to Beep? Comparing Abstract Versus
  Language-based Multimodal Driver Displays}. In
  \bibinfo{booktitle}{\emph{Proceedings of the 33rd Annual ACM Conference on
  Human Factors in Computing Systems}} (Seoul, Republic of Korea)
  \emph{(\bibinfo{series}{CHI '15})}. \bibinfo{publisher}{Association for
  Computing Machinery}, \bibinfo{address}{New York, NY, USA},
  \bibinfo{pages}{3971–3980}.
\newblock
\showISBNx{9781450331456}
\urldef\tempurl%
\url{https://doi.org/10.1145/2702123.2702167}
\showDOI{\tempurl}


\bibitem[Schmidt et~al\mbox{.}(2019)]%
        {Schmidt2019}
\bibfield{author}{\bibinfo{person}{Holger Schmidt}, \bibinfo{person}{Gottfried
  Zimmermann}, {and} \bibinfo{person}{Albrecht Schmidt}.}
  \bibinfo{year}{2019}\natexlab{}.
\newblock \showarticletitle{Using Gaze-based Interactions in Automated Vehicles
  for Increased Road Safety}. In \bibinfo{booktitle}{\emph{Proceedings of the
  11th International Conference on Automotive User Interfaces and Interactive
  Vehicular Applications: Adjunct Proceedings}} (Utrecht, Netherlands)
  \emph{(\bibinfo{series}{AutomotiveUI ’19})}.
  \bibinfo{publisher}{Association for Computing Machinery},
  \bibinfo{address}{New York, NY, USA}, \bibinfo{pages}{321–326}.
\newblock
\showISBNx{9781450369206}
\urldef\tempurl%
\url{https://doi.org/10.1145/3349263.3351910}
\showDOI{\tempurl}


\bibitem[Shah et~al\mbox{.}(2018)]%
        {shah2018airsim}
\bibfield{author}{\bibinfo{person}{Shital Shah}, \bibinfo{person}{Debadeepta
  Dey}, \bibinfo{person}{Chris Lovett}, {and} \bibinfo{person}{Ashish Kapoor}.}
  \bibinfo{year}{2018}\natexlab{}.
\newblock \showarticletitle{Airsim: High-fidelity Visual and Physical
  Simulation for Autonomous Vehicles}. In \bibinfo{booktitle}{\emph{Field and
  service robotics}}. Springer, \bibinfo{pages}{621--635}.
\newblock


\bibitem[Shahmoradi et~al\mbox{.}(2020)]%
        {Shahmoradi}
\bibfield{author}{\bibinfo{person}{Javad Shahmoradi}, \bibinfo{person}{Elaheh
  Talebi}, \bibinfo{person}{Pedram Roghanchi}, {and} \bibinfo{person}{Mostafa
  Hassanalian}.} \bibinfo{year}{2020}\natexlab{}.
\newblock \showarticletitle{A Comprehensive Review of Applications of Drone
  Technology in the Mining Industry}.
\newblock  (\bibinfo{year}{2020}).
\newblock


\bibitem[Strabala et~al\mbox{.}(2013)]%
        {Strabala2013}
\bibfield{author}{\bibinfo{person}{Kyle Strabala}, \bibinfo{person}{Min~Kyung
  Lee}, \bibinfo{person}{Anca Dragan}, \bibinfo{person}{Jodi Forlizzi},
  \bibinfo{person}{Siddhartha~S. Srinivasa}, \bibinfo{person}{Maya Cakmak},
  {and} \bibinfo{person}{Vincenzo Micelli}.} \bibinfo{year}{2013}\natexlab{}.
\newblock \showarticletitle{Toward Seamless Human-Robot Handovers}.
\newblock \bibinfo{journal}{\emph{J. Hum.-Robot Interact.}}
  \bibinfo{volume}{2}, \bibinfo{number}{1} (\bibinfo{date}{feb}
  \bibinfo{year}{2013}), \bibinfo{pages}{112–132}.
\newblock
\urldef\tempurl%
\url{https://doi.org/10.5898/JHRI.2.1.Strabala}
\showDOI{\tempurl}


\bibitem[Tanenhaus et~al\mbox{.}(1995)]%
        {Tanenhaus95}
\bibfield{author}{\bibinfo{person}{Michael~K Tanenhaus},
  \bibinfo{person}{Michael~J Spivey-Knowlton}, \bibinfo{person}{Kathleen~M
  Eberhard}, {and} \bibinfo{person}{Julie~C Sedivy}.}
  \bibinfo{year}{1995}\natexlab{}.
\newblock \showarticletitle{Integration of Visual and Linguistic Information in
  Spoken Language Comprehension}.
\newblock \bibinfo{journal}{\emph{Science}} \bibinfo{volume}{268},
  \bibinfo{number}{5217} (\bibinfo{year}{1995}), \bibinfo{pages}{1632--1634}.
\newblock


\bibitem[Tr\"{o}sterer et~al\mbox{.}(2017)]%
        {10.1145/3122986.3123020}
\bibfield{author}{\bibinfo{person}{Sandra Tr\"{o}sterer},
  \bibinfo{person}{Alexander Meschtscherjakov}, \bibinfo{person}{Alexander~G.
  Mirnig}, \bibinfo{person}{Artur Lupp}, \bibinfo{person}{Magdalena
  G\"{a}rtner}, \bibinfo{person}{Fintan McGee}, \bibinfo{person}{Rod McCall},
  \bibinfo{person}{Manfred Tscheligi}, {and} \bibinfo{person}{Thomas Engel}.}
  \bibinfo{year}{2017}\natexlab{}.
\newblock \showarticletitle{What We Can Learn from Pilots for Handovers and
  (de)skilling in Semi-autonomous Driving: An Interview Study}. In
  \bibinfo{booktitle}{\emph{Proceedings of the 9th International Conference on
  Automotive User Interfaces and Interactive Vehicular Applications}}
  (Oldenburg, Germany) \emph{(\bibinfo{series}{AutomotiveUI ’17})}.
  \bibinfo{publisher}{Association for Computing Machinery},
  \bibinfo{address}{New York, NY, USA}, \bibinfo{pages}{173–182}.
\newblock
\showISBNx{9781450351508}
\urldef\tempurl%
\url{https://doi.org/10.1145/3122986.3123020}
\showDOI{\tempurl}


\bibitem[van~der Heiden et~al\mbox{.}(2017a)]%
        {Heiden2017}
\bibfield{author}{\bibinfo{person}{Remo~M.A. van~der Heiden},
  \bibinfo{person}{Shamsi~T. Iqbal}, {and} \bibinfo{person}{Christian~P.
  Janssen}.} \bibinfo{year}{2017}\natexlab{a}.
\newblock \showarticletitle{Priming Drivers before Handover in Semi-autonomous
  Cars}. In \bibinfo{booktitle}{\emph{Proceedings of the 2017 CHI Conference on
  Human Factors in Computing Systems}} (Denver, Colorado, USA)
  \emph{(\bibinfo{series}{CHI '17})}. \bibinfo{publisher}{Association for
  Computing Machinery}, \bibinfo{address}{New York, NY, USA},
  \bibinfo{pages}{392–404}.
\newblock
\showISBNx{9781450346559}
\urldef\tempurl%
\url{https://doi.org/10.1145/3025453.3025507}
\showDOI{\tempurl}


\bibitem[van~der Heiden et~al\mbox{.}(2017b)]%
        {vanderHeiden}
\bibfield{author}{\bibinfo{person}{Remo~M.A. van~der Heiden},
  \bibinfo{person}{Shamsi~T. Iqbal}, {and} \bibinfo{person}{Christian~P.
  Janssen}.} \bibinfo{year}{2017}\natexlab{b}.
\newblock \showarticletitle{Priming Drivers before Handover in Semi-autonomous
  Cars}. In \bibinfo{booktitle}{\emph{Proceedings of the 2017 CHI Conference on
  Human Factors in Computing Systems}} (Denver, Colorado, USA)
  \emph{(\bibinfo{series}{CHI '17})}. \bibinfo{publisher}{Association for
  Computing Machinery}, \bibinfo{address}{New York, NY, USA},
  \bibinfo{pages}{392–404}.
\newblock
\showISBNx{9781450346559}
\urldef\tempurl%
\url{https://doi.org/10.1145/3025453.3025507}
\showDOI{\tempurl}


\bibitem[Veitengruber(1978)]%
        {Veitengruber1978}
\bibfield{author}{\bibinfo{person}{J.E. Veitengruber}.}
  \bibinfo{year}{1978}\natexlab{}.
\newblock \showarticletitle{Design Criteria for Aircraft Warning, Caution, and
  Advisory Alerting Systems}.
\newblock \bibinfo{journal}{\emph{Journal of Aircraft}} \bibinfo{volume}{15},
  \bibinfo{number}{9} (\bibinfo{year}{1978}), \bibinfo{pages}{574--581}.
\newblock
\urldef\tempurl%
\url{https://doi.org/10.2514/3.58409}
\showDOI{\tempurl}


\bibitem[Walch et~al\mbox{.}(2015)]%
        {Walch}
\bibfield{author}{\bibinfo{person}{Marcel Walch}, \bibinfo{person}{Kristin
  Lange}, \bibinfo{person}{Martin Baumann}, {and} \bibinfo{person}{Michael
  Weber}.} \bibinfo{year}{2015}\natexlab{}.
\newblock \showarticletitle{Autonomous Driving: Investigating the Feasibility
  of Car-driver Handover Assistance}. In \bibinfo{booktitle}{\emph{Proceedings
  of the 7th International Conference on Automotive User Interfaces and
  Interactive Vehicular Applications}} (Nottingham, United Kingdom)
  \emph{(\bibinfo{series}{AutomotiveUI '15})}. \bibinfo{publisher}{Association
  for Computing Machinery}, \bibinfo{address}{New York, NY, USA},
  \bibinfo{pages}{11–18}.
\newblock
\showISBNx{9781450337366}
\urldef\tempurl%
\url{https://doi.org/10.1145/2799250.2799268}
\showDOI{\tempurl}


\bibitem[Walch et~al\mbox{.}(2019)]%
        {Walch2019}
\bibfield{author}{\bibinfo{person}{Marcel Walch}, \bibinfo{person}{David Lehr},
  \bibinfo{person}{Mark Colley}, {and} \bibinfo{person}{Michael Weber}.}
  \bibinfo{year}{2019}\natexlab{}.
\newblock \showarticletitle{Don’t You See Them? Towards Gaze-based
  Interaction Adaptation for Driver-vehicle Cooperation}. In
  \bibinfo{booktitle}{\emph{Proceedings of the 11th International Conference on
  Automotive User Interfaces and Interactive Vehicular Applications: Adjunct
  Proceedings}} (Utrecht, Netherlands) \emph{(\bibinfo{series}{AutomotiveUI
  ’19})}. \bibinfo{publisher}{Association for Computing Machinery},
  \bibinfo{address}{New York, NY, USA}, \bibinfo{pages}{232–237}.
\newblock
\showISBNx{9781450369206}
\urldef\tempurl%
\url{https://doi.org/10.1145/3349263.3351338}
\showDOI{\tempurl}


\bibitem[Wasinger et~al\mbox{.}(2005)]%
        {E2:wasinger2005integrating}
\bibfield{author}{\bibinfo{person}{Rainer Wasinger}, \bibinfo{person}{Antonio
  Kr{\"u}ger}, {and} \bibinfo{person}{Oliver Jacobs}.}
  \bibinfo{year}{2005}\natexlab{}.
\newblock \showarticletitle{Integrating Intra and Extra Gestures into a Mobile
  and Multimodal Shopping Assistant}. In
  \bibinfo{booktitle}{\emph{International Conference on Pervasive Computing}}.
  \bibinfo{pages}{297--314}.
\newblock
\urldef\tempurl%
\url{https://doi.org/10.1007/11428572_18}
\showDOI{\tempurl}


\bibitem[Wiehr et~al\mbox{.}(2021)]%
        {Wiehr2021}
\bibfield{author}{\bibinfo{person}{Frederik Wiehr}, \bibinfo{person}{Anke
  Hirsch}, \bibinfo{person}{Lukas Schmitz}, \bibinfo{person}{Nina Knieriemen},
  \bibinfo{person}{Antonio Kr\"{u}ger}, \bibinfo{person}{Alisa Kovtunova},
  \bibinfo{person}{Stefan Borgwardt}, \bibinfo{person}{Ernie Chang},
  \bibinfo{person}{Vera Demberg}, \bibinfo{person}{Marcel Steinmetz}, {and}
  \bibinfo{person}{J\"{o}rg Hoffmann}.} \bibinfo{year}{2021}\natexlab{}.
\newblock \bibinfo{booktitle}{\emph{Why Do I Have to Take Over Control?
  Evaluating Safe Handovers with Advance Notice and Explanations in Had}}.
\newblock \bibinfo{publisher}{Association for Computing Machinery},
  \bibinfo{address}{New York, NY, USA}, \bibinfo{pages}{308–317}.
\newblock
\showISBNx{9781450384810}
\urldef\tempurl%
\url{https://doi.org/10.1145/3462244.3479884}
\showURL{%
\tempurl}


\bibitem[{Wiehr} et~al\mbox{.}(2021)]%
        {WiehrICMI2021}
\bibfield{author}{\bibinfo{person}{Frederik {Wiehr}}, \bibinfo{person}{Anke
  {Hirsch}}, \bibinfo{person}{Lukas {Schmitz}}, \bibinfo{person}{Nina
  {Knieriemen}}, \bibinfo{person}{Antonio {Krüger}}, \bibinfo{person}{Alisa
  {Kovtunova}}, \bibinfo{person}{Stefan {Borgwardt}}, \bibinfo{person}{Ernie
  {Chang}}, \bibinfo{person}{Vera {Demberg}}, \bibinfo{person}{Marcel
  {Steinmetz}}, {and} \bibinfo{person}{Jörg {Hoffmann}}.}
  \bibinfo{year}{2021}\natexlab{}.
\newblock \showarticletitle{Why Do {I} Have to Take Over Control? {E}valuating
  Safe Handovers with Advance Notice and Explanations in {HAD}}. In
  \bibinfo{booktitle}{\emph{{ICMI}'21: International Conference on Multimodal
  Interaction}}, \bibfield{editor}{\bibinfo{person}{Zakia {Hammal}},
  \bibinfo{person}{Carlos {Busso}}, \bibinfo{person}{Catherine {Pelachaud}},
  \bibinfo{person}{Sharon~L. {Oviatt}}, \bibinfo{person}{Albert~Ali {Salah}},
  {and} \bibinfo{person}{Guoying {Zhao}}} (Eds.). \bibinfo{publisher}{{ACM}},
  \bibinfo{pages}{308--317}.
\newblock
\urldef\tempurl%
\url{https://doi.org/10.1145/3462244.3479884}
\showDOI{\tempurl}


\end{thebibliography}

\end{document}